\documentclass[aps,prl,twocolumn,superscriptaddress,longbibliography]{revtex4-1}
\pdfoutput=1
\usepackage{amsmath}
\usepackage{epsfig}
\usepackage{graphicx}
\usepackage{graphicx, color, epstopdf}
\usepackage{bm}
\usepackage{amssymb}
\usepackage{hyperref}
\usepackage{xcolor}
\usepackage{subfigure}
\usepackage{amsmath}
\usepackage{caption}
\usepackage{graphicx}
\usepackage{subfigure}
\usepackage{bm}
\usepackage{amssymb}
\usepackage{hyperref}
\usepackage{xcolor}
\usepackage{chngcntr}
\usepackage{titlesec}

\begin{document}

\title{Cyclotron quantization and mirror-time transition on nonreciprocal lattices}
\author{Kai Shao}
\affiliation{National Laboratory of Solid State Microstructures, School of Physics,
and Collaborative Innovation Center of Advanced Microstructures, Nanjing University, Nanjing 210093, China}

\author{Zhuo-Ting Cai}
\affiliation{National Laboratory of Solid State Microstructures, School of Physics,
and Collaborative Innovation Center of Advanced Microstructures, Nanjing University, Nanjing 210093, China}

\author{Hao Geng}
\affiliation{National Laboratory of Solid State Microstructures, School of Physics,
and Collaborative Innovation Center of Advanced Microstructures, Nanjing University, Nanjing 210093, China}

\author{Wei Chen}
\email{Corresponding author: pchenweis@gmail.com}
\affiliation{National Laboratory of Solid State Microstructures, School of Physics,
and Collaborative Innovation Center of Advanced Microstructures, Nanjing University, Nanjing 210093, China}

\author{D. Y. Xing}
\affiliation{National Laboratory of Solid State Microstructures, School of Physics,
and Collaborative Innovation Center of Advanced Microstructures, Nanjing University, Nanjing 210093, China}
\begin{abstract}
Unidirectional transport and localized cyclotron
motion are two opposite physical phenomena.
Here, we study the interplay effects between them on nonreciprocal
lattices subject to a magnetic field.
We show that, in the long-wavelength limit, the trajectories of the wave packets
always form closed orbits in four-dimensional (4D)
complex space. Therefore,
the semiclassical quantization rules
persist despite the nonreciprocity, which preserves real Landau levels.
We predict a different type of non-Hermitian spectral transition
induced by the spontaneous breaking
of the combined mirror-time reversal ($\mathcal{MT}$) symmetry,
which generally exists in such systems. An order parameter is proposed to
describe the $\mathcal{MT}$ phase transition, not only to
determine the $\mathcal{MT}$ phase boundary
but also to quantify the degree of $\mathcal{MT}$-symmetry breaking.
Such an order parameter can be generally
applied to all types of non-Hermitian phase transitions.
\end{abstract}

\date{\today}

\maketitle
\emph{Introduction.-}Non-Hermitian physics~\cite{Bender98prl,bender2007making,ashida2020non}
has attracted growing research interest recently
for its intriguing
properties and potential applications
that can be implemented in various physical systems, including photonic systems~\cite{feng2017non,Ozawa2019rmp,chen2017exceptional,el2018non,ozdemir2019parity,
miri2019exceptional},
open quantum systems coupled to the environment~\cite{rotter1991continuum,rotter2009non,
Malzard2015prl,diehl2008quantum}, quasiparticles in condensed matter~\cite{kozii2017nonhermitian,shen2018prl,Papaj2019prb,Yoshida18prb},
and electrical
circuits~\cite{Schindler2011pra,lee2018topolectrical,kotwal2021active,Ningyuan15prx,imhof2018topolectrical}.
The non-Hermitian topological band theory has
been studied extensively and achieved
plentiful results~\cite{shen18prl2,Bergholtz21rmp}, such as
anomalous edge modes~\cite{Lee2016prl,xiong2018does},
enriched topological phases~\cite{Gong2018prx,kawabata19nc,Xu2017prl,Carlstr19prb},
and topological lasing~\cite{peng2014parity,st2017lasing,Parto18prl}.
It is now well accepted that the conventional
Bloch band theory
should be replaced by the non-Bloch band theory
for non-Hermitian systems
with the so-called non-Hermitian skin effect (NHSE)~\cite{Yaoshunyu2018prl,Kunst2018prl}.

\begin{figure*}
\centering
\includegraphics[width=1.7\columnwidth]{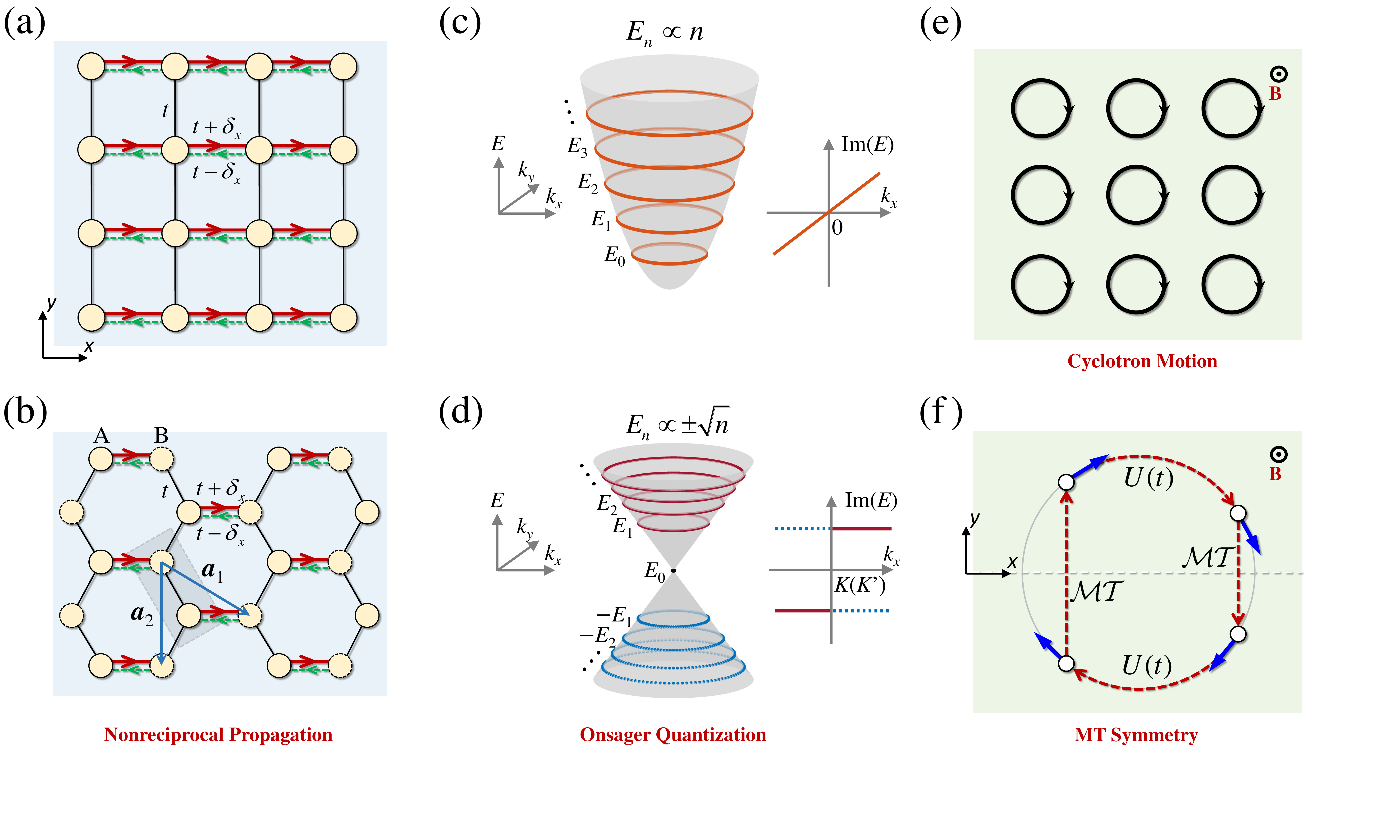}\\
\caption{Schematic illustration of
(a) nonreciprocal Harper-Hofstadter model and (b) nonreciprocal honeycomb lattice model
with unequal hopping strengths
$t\pm\delta_x$ in the $x$-direction
and equal hopping $t$ in others.
The gray dashed box in (b) denotes the unit cell composed of two sites $A$, $B$ and $\bm{a}_{1,2}$ are
the unit vectors.
(c) Low-energy parabolic dispersion with the linear imaginary part
corresponding to the square lattice model in (a).
(d) Dirac cone dispersion and its imaginary part corresponding to the honeycomb lattice in (b).
The signs of the imaginary part of the energy coincide with those of the velocity.
The contours in (c) and (d) are the closed orbits that satisfy
the Onsager-Lifshitz quantization rule.
(e) Semiclassical cyclotron motion of charged particles in
a magnetic field. (f) Semiclassical picture
of the $\mathcal{MT}$ symmetry, in which
the successive actions of the $\mathcal{MT}$ operation ($y\rightarrow-y, v_x\rightarrow-v_x$) and the time evolution
$U(t)$ leave the state unchanged.}\label{fig1}
\end{figure*}

The NHSE is a unique phenomenon~\cite{Yaoshunyu2018prl,Kunst2018prl,Yokomizo2019prl,Borgnia20prl,
Borgnia20prl,Zhangkai2020prl,Okuma2020prl,Yangzhesen2020prl,li2020critical}
meaning that all the bulk states are
driven to the edge of the system under the open boundary condition (OBC),
which has been confirmed in recent experiments in
various physical systems~\cite{helbig2020generalized,xiao2020non,weidemann2020topological,
ghatak2020observation,palacios2021guided,zhang2021observation,liang2022observation}.
Its intriguing interplay with the parity-time ($\mathcal{PT}$) phase transition
has attracted attentions very recently~\cite{longhi2019non,Longhi19prr,xiao21prl,song2021non},
which opened up the possibility of manipulating the $\mathcal{PT}$ transition by the NHSE.
Physically, the NHSE in one-dimensional (1D) systems originates from
the point gap topology of the
energy spectra under the periodic boundary condition (PBC)~\cite{Borgnia20prl,Okuma2020prl,
Zhangkai2020prl}, which is manifested as
the nonreciprocal propagation of the wave packet through the system~\cite{Yi20prl,Xue21prb}.
Such nonreciprocity as shown in Figs.~\ref{fig1}(a) and (b) is
a particular type of delocalization effect and can induce
a delocalization transition~\cite{HN96prl,Chenshu19prb,
Chenshu21prb}.

In contrast to nonreciprocal propagation,
a magnetic field in a 2D system leads to the opposite effect.
The motion of charged particles in a
magnetic field forms cyclotron
orbits with the guiding centers localized in space;
see Fig.~\ref{fig1}(e). The quantization of these cyclotron orbits
results in flat Landau bands with zero mobility, which is incompatible with
the picture of nonreciprocal propagation. Given that a magnetic field and
nonreciprocity may coexist
in a variety of natural and artificial systems~\cite{abo2001observation,lin2009synthetic,Zhang20prl,
lin2021steeringOL}, it is interesting to explore
their fascinating interplay and the resultant
physical effects. Open questions that naturally arise
include the robustness of cyclotron
orbits as well as their quantization against nonreciprocity
and possible new types of non-Hermitian phase transitions, etc.

In this Letter, we study the physical effects
in nonreciprocal systems
subject to a magnetic field.
We show that semiclassical trajectories of
the wave packets always form closed orbits in
the 4D complex space
in the long-wavelength limit despite
the nonreciprocity.
As a result, the Onsager-Lifshitz quantization rule
persists, which protects real Landau levels from being complex.
Moreover, we show that such non-Hermitian magnetic systems
generally possess an
inherent mirror-time reversal ($\mathcal{MT}$)
symmetry, which dictates
a spectral phase transition,
dubbed the $\mathcal{MT}$ transition.
Specifically, a real-to-complex spectral transition
occurs along with the spontaneous breaking of
the $\mathcal{MT}$ symmetry.
An order parameter is proposed to quantify
the $\mathcal{MT}$-symmetry breaking,
which not only gives a definite phase boundary but also
specifies to what extent the symmetry is broken.
Our work generalizes the celebrated $\mathcal{PT}$
physics~\cite{el2018non,ozdemir2019parity,miri2019exceptional}
to the $\mathcal{MT}$ scenario in a class of magnetic systems,
which may lead to interesting observations and
applications.

\emph{Model.-}To be concrete, we first study
the nonreciprocal square lattice in Fig.~\ref{fig1}(a) and then
verify the universality of the results
on the honeycomb lattice in Fig.~\ref{fig1}(b).
A square lattice with nonreciprocal hopping
under a magnetic field $B$
can be described by the modified Harper-Hofstadter model as~\cite{harper1955single,Hofstadter1976prb}
\begin{equation}\label{H}
\begin{split}
H&=-\sum_{m,n}\big(t_x^+ c^{\dagger}_{m+1,n}c_{m,n}+t_x^- c^{\dagger}_{m,n}c_{m+1,n}\\
&+t e^{i2\pi m\phi}c^{\dagger}_{m,n+1}c_{m,n}+t e^{-i2\pi m\phi}c^{\dagger}_{m,n}c_{m,n+1}\big),\\
\end{split}
\end{equation}
where $c^{\dagger}_{m,n}$ ($c_{m,n}$) are the creation (annihilation)
operator on the site $(m,n)$,
$t_{x}^\pm=t\pm\delta_{x}$ describe the nonreciprocal hopping
in the $x$-direction
with $\delta_{x}$ the strength of nonreciprocity.
The phase factor $\phi=\Phi/\Phi_0$
is defined by the magnetic flux
$\Phi=Ba^2$ through a lattice cell (lattice constant $a$) divided by
the flux quantum $\Phi_0=h/q$ with $q$ the charge of the particle.
Here, the Landau gauge $\bm{A}=(0, Bx)$ has been adopted.
In the rest of this Letter, we set $h=q=a=1$ in all
numerical calculations for simplicity
and denote the OBC and PBC
in the $\alpha$-direction ($\alpha=x,y$) as
$\alpha$-OBC and $\alpha$-PBC for brevity.

For $B=0$, the energy spectrum under the $x,y$-PBC
is $E(\bm{k})=-2t(\cos k_x+\cos k_y)+2i\delta_x\sin k_x$
with $\bm{k}=(k_{x}, k_y)$ the wave vector.
The low-energy expansion at the band bottom
yields the parabolic dispersion plus an imaginary part
as $\varepsilon(\bm{k})= t(k_x^2+k_y^2)+2i\delta_x k_x$; see Fig.~\ref{fig1}(c),
which resembles a non-Hermitian normal particle.
The odd function with $\text{Im}[E(k_x)]=-\text{Im}[E(-k_x)]$
induces a point gap topology for each
transverse $k_y$-mode, which
results in the nonreciprocal propagation of the wave packet in the $x$-direction~\cite{Yi20prl,Xue21prb}.
Accordingly, the system exhibits NHSE in the $x$-direction
under the $x$-OBC~\cite{Borgnia20prl,Okuma2020prl,
Zhangkai2020prl}, which can be read from the right eigenfunctions of Eq.~\eqref{H} under the $x,y$-OBC as
$\psi^R(x,y)=\psi_{m,n}=(t_x^+/t_x^-)^{m/2}\sin{(m k_x)}\sin{(n k_y)}$.
A positive $\delta_{x}$ results in an envelope function $(t_x^+/t_x^-)^{m/2}$
on top of the standing waves so that all the wave functions
are localized at the right boundary,
namely NHSE. Due to its incompatibility, a small magnetic field is sufficient to
drive the skin modes to penetrate deeply into the bulk,
showing a considerable suppression of the NHSE~\cite{sm,lu2021magnetic}.
Physically, it stems from the shrinkage of the point gap
for each $k_y$ channel, i.e., a reduction of
the nonreciprocity~\cite{sm}.

\begin{figure}[ht]
\centering
\includegraphics[width=1\columnwidth]{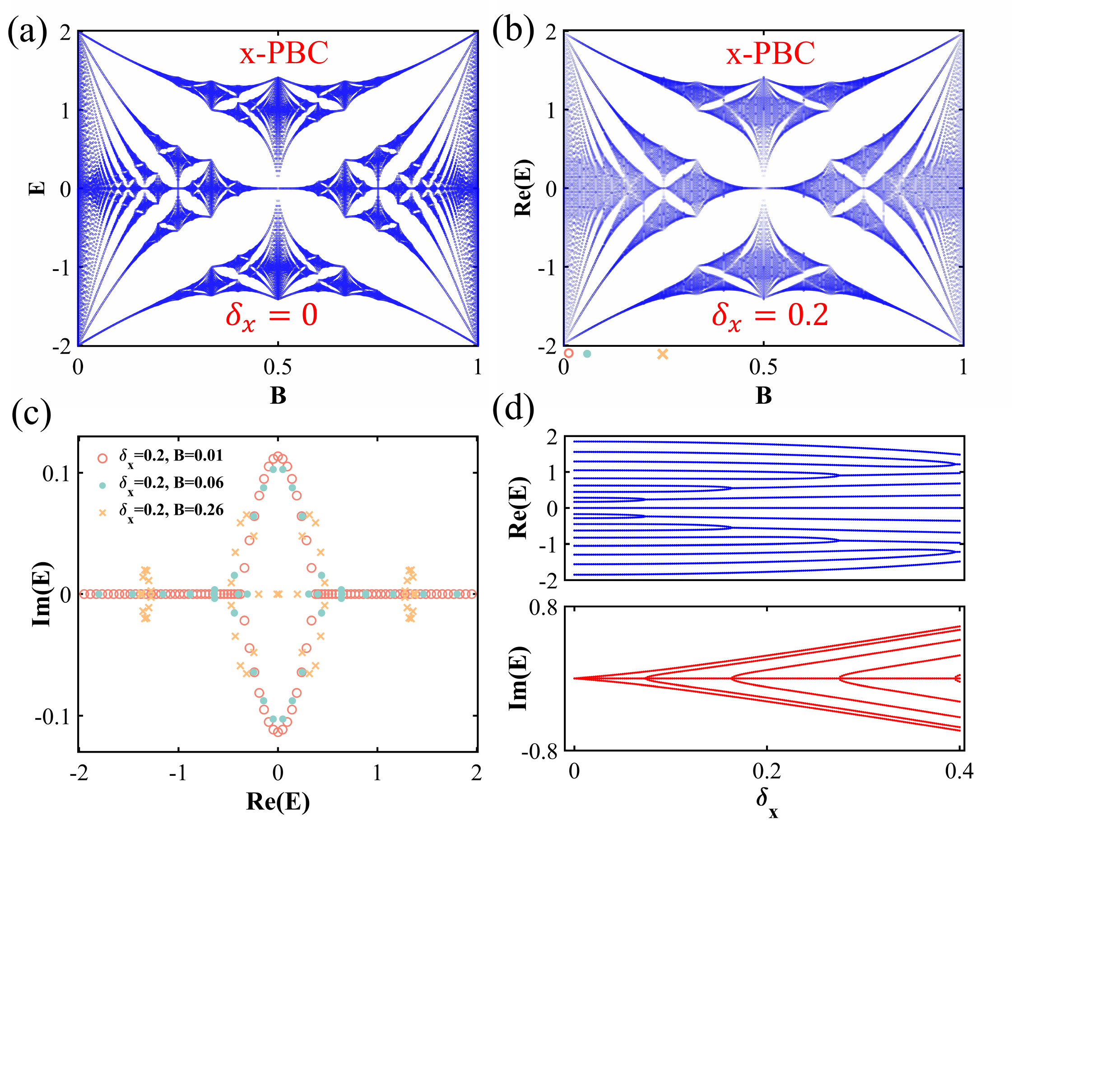}\\
\caption{Energy spectra under
the $x,y$-PBC calculated in the momentum space
for (a) $\delta_x=0$ and (b) $\delta_x=0.2$.
(c) Complex energy spectra for different $B$ marked
in (b). (d) Complex
energy spectra as a function of $\delta_x$ with $B=0.05$.
In all figures, $t=0.5$.}
\label{fig2}
\end{figure}

\emph{Semiclassical Onsager-Lifshitz
quantization.}-It is of particular interest
to investigate the semiclassical quantization
of magnetic cyclotron orbits
subject to nonreciprocal
propagation. Intuitively, nonreciprocity
tends to break those closed orbits [cf. Fig.~\ref{fig1}(e)]
and thus the quantization condition.
To study the wave packet dynamics inside the bulk
and get rid of the boundary effects, we
adopt the $x,y$-PBC.
The Hamiltonian~\eqref{H} can be written
in momentum space and diagonalized directly, with
the energy spectra for $\delta_{x}=0$ and $\delta_x>0$
shown in Figs.~\ref{fig2}(a) and \ref{fig2}(b), respectively.

For $\delta_{x}=0$, Fig.~\ref{fig2}(a) presents
the familiar butterfly diagram.
The Landau fan structure near the
band edges for small $B$ indicates
the high degeneracy of Landau levels
with vanishing band width,
which stems from localized cyclotron motion.
A finite nonreciprocal hopping $\delta_x$ leads to visible modifications
in the energy spectra with
complex energy spectra showing up
at the band center;
see Figs.~\ref{fig2}(b) and (c).
As a result, the self-similar fractal patterns merge into
continuous pieces along with multiple gap closings.
Notably, as $\delta_x$ increases, the energy levels
coalesce in pairs and create multiple exceptional points;
see Fig.~\ref{fig2}(d). Given the high degeneracy of the
magnetic spectra, a large number of exceptional points
can be implemented in such systems.

Remarkably, one can find that the Landau levels near the band top and bottom
remain unchanged despite the nonreciprocity
by comparing Figs.~\ref{fig2}(a)
and \ref{fig2}(b).
Moreover, these low-energy Landau levels remain real [see Fig.~\ref{fig2}(c)],
which indicates that
the magnetic field prevents the system from a
real-to-complex spectral transition
in the long-wavelength limit. It can be shown that the
quantized energy levels exhibit the scaling $E_n\propto nB$~\cite{sm},
which reduces to the behavior of free particles with a quadratic
dispersion [cf. Fig.~\ref{fig1}(c)]. Such observations indicate that
the Onsager-Lifshitz quantization rule~\cite{onsager1952interpretation,lifshitz1954}
remains valid even in the presence of nonreciprocal propagation.

\begin{figure}[!htbp]
\centering
\includegraphics[width=1\columnwidth]{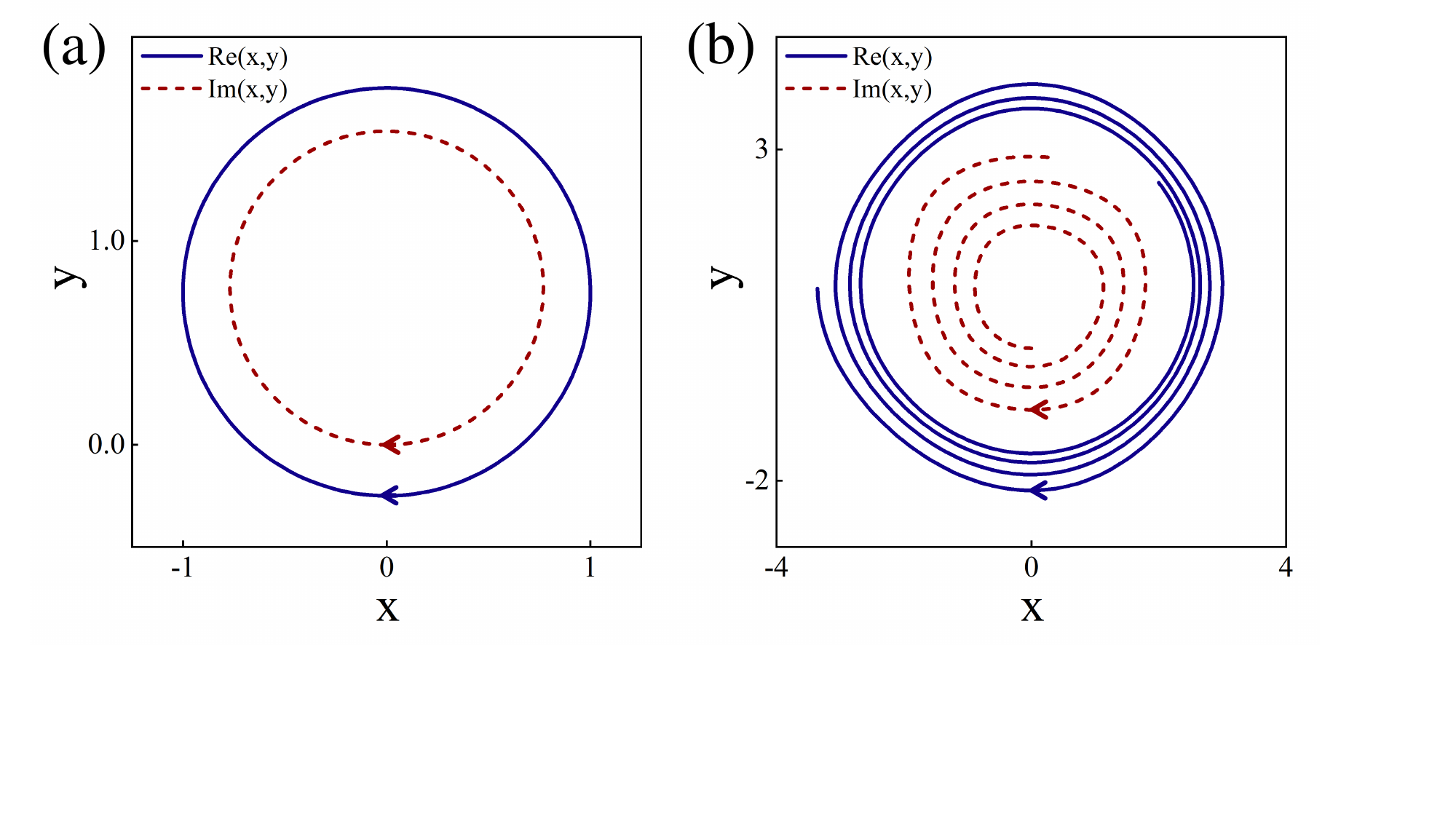}\\
\caption{Projections of the semiclassical trajectories in the real
and imaginary $x$-$y$ planes. (a) Closed orbits in the
long-wavelength limit with $x(0)=1$,
$y(0)=0.75$, $p_{x}(0)=0$, $p_{y}(0)=0$ and (b) open trajectories beyond this limit with $x(0)=2$,
$y(0)=2.5$, $p_{x}(0)=0.4$, $p_{y}(0)=0$. Other parameters are $B=0.26$,
$\delta_x=0.1$.  }\label{fig3}
\end{figure}

We prove this assertion based on the semiclassical equation of motion.
In the long-wavelength limit, the effective Hamiltonian
in a magnetic field can be written as
$
\varepsilon(\bm{p}-q\bm{A})=t(\bm{p}-q\bm{A})^2+2i\delta_x k_x
$
through Peierls substitution $\hbar\bm{k}\rightarrow\bm{p}-q\bm{A}$ with
$\bm{p}$ the canonical momentum. By solving the canonical equation
$\dot{\bm{p}}=-\nabla_{\bm{r}}\varepsilon, \dot{\bm{r}}=\nabla_{\bm{p}}\varepsilon,
$ we obtain the time($\tau$)-dependent coordinate functions as~\cite{sm}
\begin{equation}
\begin{split}
x(\tau)&=A_1 e^{i\omega\tau}+A_2e^{-i\omega\tau}+x_0,\\
y(\tau)&=i(A_1 e^{i\omega\tau}-A_2e^{-i\omega\tau})+y_0,
\end{split}
\end{equation}
with the parameters $A_{1,2}, x_0, y_0$ determined
by the initial conditions and $\omega=2Btq/\hbar^2$ the cyclotron frequency.
The solutions clearly show that the semiclassical
trajectories of the wave packet always
form closed orbits despite the nonreciprocity; see Fig.~\ref{fig3}(a).
However, different from the Hermitian case, the closed orbits
generally reside in 4D complex $x$-$y$ space~\cite{bender2007making}.

Although closed orbits are formed in higher-dimensional coordinate space,
the single-valuedness of the
wave function for the periodic motion
still imposes the quantization condition,
which is the non-Hermitian Onsager-Lifshitz quantization rule~\cite{onsager1952interpretation,lifshitz1954}:
\begin{equation}
\oint \bm{p}\cdot d \bm{r}=(n+\frac{1}{2})h.
\end{equation}
Substituting the solutions $\bm{p}(\tau), \bm{r}(\tau)$ into
the quantization condition results in exactly the
Landau levels $\varepsilon_n=(n+1/2)\hbar\omega+\delta_x^2/t$
of a normal particle apart from a small shift $\delta_x^2/t$~\cite{sm}, consistent with the
numerical results in Fig.~\ref{fig2}(b).
Therefore, we conclude that
the semiclassical quantization is robust
against the nonreciprocity and thus can
protect the energy from being complex.
Meanwhile, as $E(\bm{k})$ deviates from
the long-wavelength limit, the coordinate
functions can be solved numerically, which exhibit
unclosed trajectories in the complex $x$-$y$ space; see Fig.~\ref{fig3}(b).
As a result, the energy spectra become complex due to the nonreciprocal
hopping as those in Figs.~\ref{fig2}(b) and (c).

%
%
%

\emph{$\mathcal{MT}$ phase transition.}-The properties of the whole energy spectra
strongly depend on the boundary conditions in the $x$-direction.
The magnetic spectra are entirely real for the $x$-OBC~\cite{sm}
and partially complex for the $x$-PBC.
We will show that a real-to-complex transition
of the entire spectra can be implemented,
which is associated with the spontaneous breaking of
the inherent $\mathcal{MT}$ symmetry of the systems.

The Hamiltonian~\eqref{H} possesses the combined
$\mathcal{MT}$ symmetry as
\begin{equation}\label{mt}
\mathcal{M}\mathcal{T}H(\mathcal{M}\mathcal{T})^{-1}=H,
\end{equation}
with the operators of mirror reflection ($\mathcal{M}$) about the $x$-axis
and time reversal ($\mathcal{T}$) defined by
\begin{equation}
\mathcal{M}c_{m,n}\mathcal{M}^{-1}=c_{m,-n},\mathcal{T}c_{m,n}\mathcal{T}^{-1}=c_{m,n}, \mathcal{T}i\mathcal{T}^{-1}=-i.
\end{equation}
The $\mathcal{MT}$ symmetry can be understood by the semiclassical picture as
illustrated in Fig.~\ref{fig1}(f). A quantum state
under successive actions of the $\mathcal{MT}$
operation and the time evolution $U(t)$ remains the same,
i.e., $U(t)\mathcal{MT}U(t)\mathcal{MT}=1$.
The constraint by the $\mathcal{MT}$ symmetry
can be rewritten in
another standard form as
$
\mathcal{MO}H^\dag(\mathcal{MO})^{-1}=H,
$
with $\mathcal{O}$ the transpose operation,
and then $H$ is said to be $\mathcal{MO}$-pseudo-Hermition~\cite{ashida2020non}.
As a result, the energy spectra can be either entirely real
or composed of complex conjugate pairs.
For a specific state,
the real (complex) nature of the energy
corresponds to its wave function with
(without) the $\mathcal{MT}$ symmetry~\cite{wigner1993normal}.

The $\mathcal{MT}$-symmetry breaking
for the $i$th right eigenstate $\psi^R_i$ can be
measured by the Hilbert-Schmidt
quantum distance $d^i_{\text{HS}}$~\cite{shapere1989geometric}, which is given by
\begin{equation}
d_{\text{HS}}^i=\sqrt{1-|\langle\mathcal{MT}\rangle_i|^2},\ \langle\mathcal{MT}\rangle_i=\langle\psi_i^R|\mathcal{MT}|\psi_i^R\rangle.
\end{equation}
It characterizes the quantum mechanical distance between
the wave functions before and after the $\mathcal{MT}$ operation.
For the state that satisfies the $\mathcal{MT}$ symmetry,
the $\mathcal{MT}$ operation yields only an overall phase factor,
i.e., $\mathcal{MT}|\psi_i^R\rangle=e^{i\theta}|\psi_i^R\rangle$,
so that $d_{\text{HS}}^i=0$.
In contrast, if the state breaks the $\mathcal{MT}$ symmetry,
one has $0<d_{\text{HS}}^i\leq1$.

It is convenient to introduce
an \emph{order parameter} to quantify
the spontaneous symmetry breaking in non-Hermitian
phase transitions.
An insightful choice of the order parameter can
be the average quantum distance
of all $\mathcal{N}$ eigenstates defined as
\begin{equation}\label{d}
d_{\text{HS}}=\frac{1}{\mathcal{N}}\sum_i^{\mathcal{N}} d^i_{\text{HS}}.
\end{equation}
The $\mathcal{MT}$-symmetric and $\mathcal{MT}$-broken
phases correspond to $d_{\text{HS}}=0$ and $d_{\text{HS}}>0$, respectively,
resembling the spontaneous symmetry breaking in
continuous phase transitions.
Importantly, in addition to being a criterion of the $\mathcal{MT}$ transition, the magnitude of
$d_{\text{HS}}$ can tell to what extent the $\mathcal{MT}$ symmetry is broken.

A tunable boundary condition~\cite{xiong2018does,Guo21prl}
that can drive a continuous $\mathcal{MT}$ transition
is defined by the boundary hopping
$-\gamma_B(t_x^+ c^{\dagger}_{1,n}c_{M,n}+t_x^- c^{\dagger}_{M,n}c_{1,n})$.
The parameter $\gamma_B\in[0,1]$ and its two limits
$\gamma_B=0$ and $\gamma_B=1$ correspond to
the $x$-OBC and $x$-PBC, respectively.
We perform Fourier transformation to the bulk Hamiltonian
in the $y$-direction and rewrite it as $\tilde{H}=-\sum_{m,k_y}\big[t_x^+c^{\dagger}_{m+1,k_y}c_{m,k_y}+t_x^-c^{\dagger}_{m,k_y}c_{m+1,k_y}
+2t\cos {(k_y+2\pi m \phi)}c^{\dagger}_{m,k_y}c_{m,k_y}\big]$.
Its eigenstates $\psi^R_i(m,k_y)$ are labeled by $i$ and $k_y$.
In this representation, it can be proved that the $\mathcal{MT}$ operator acts on the
wave function as $\mathcal{MT}\psi^R_i(m,k_y)=\psi^{R*}_i(m,k_y)$.
The order parameter
$d_{\text{HS}}$ as a function of $\gamma_B$ and $\delta_x$
can be calculated by Eq.~\eqref{d}, in which the average is taken over
all states labeled by $i$ and $k_y$.

\begin{figure}[!htbp]
\centering
\includegraphics[width=0.98\columnwidth]{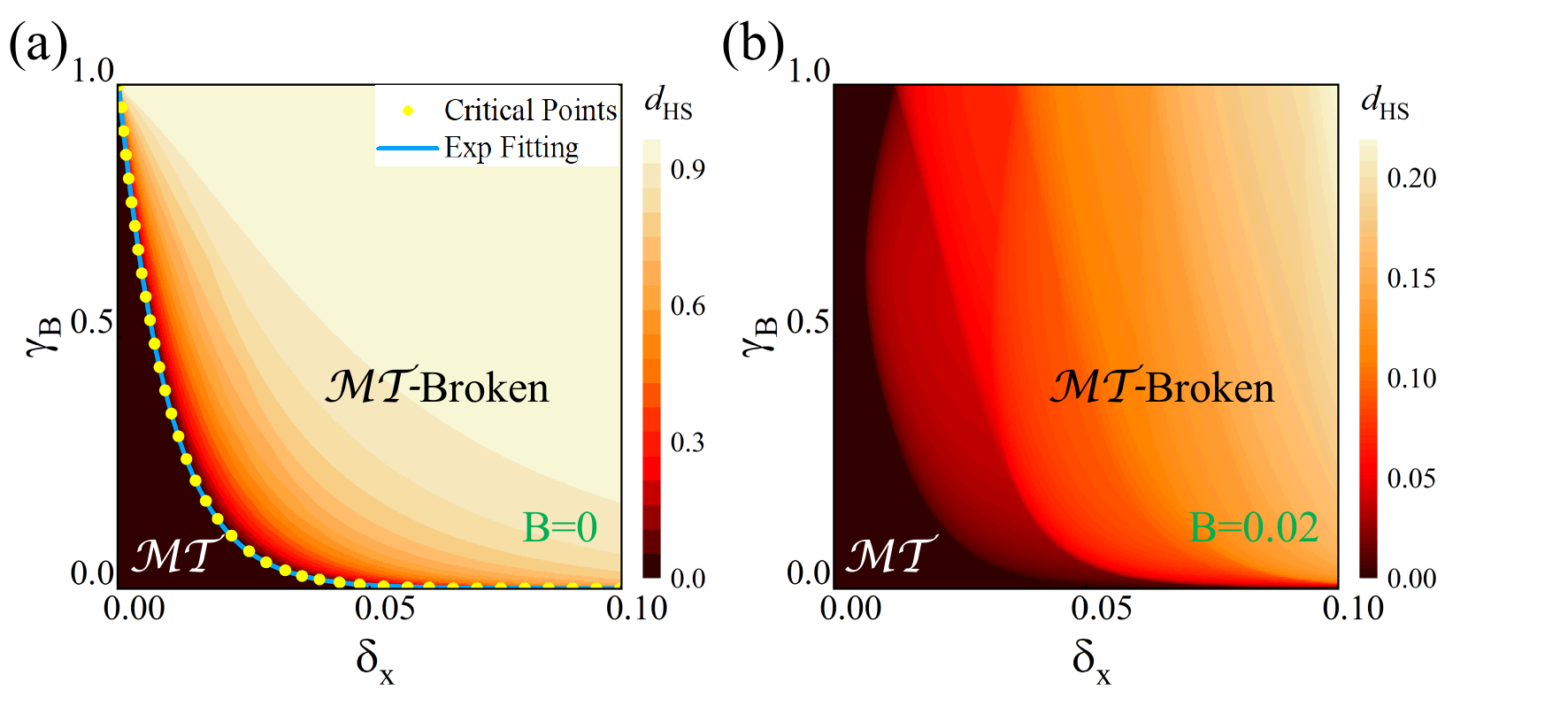}\\
\caption{Order parameter $d_{\text{HS}}$ as
a function of $\delta_x$ and $\gamma_B$ for the nonreciprocal square lattice with (a) $B=0$ and (b) $B=0.02$.
Critical points in (a) mark the real-to-complex
spectral transition and the phase boundary defined by
the $d_{\text{HS}}$ contours is fitted by exponential functions.
Other parameters are $M=50$, $t=0.5$.}\label{fig4}
\end{figure}

Numerical results of $d_{\text{HS}}$ for zero and finite $B$
are shown in Fig.~\ref{fig4}.
One can see that there is a clear phase boundary formed
between the $\mathcal{MT}$-symmetric ($d_{\text{HS}}=0$)
and $\mathcal{MT}$-broken ($d_{\text{HS}}>0$)
regions. Such a phase boundary can also be
obtained by the critical points of
the real-to-complex spectral
transition as usually done in the literature, and
the phase boundaries obtained by the two methods
show good coincidence; see Fig.~\ref{fig4}(a).
This is assured by
the theorem associated with the $\mathcal{MT}$
antiunitary symmetry~\cite{wigner1993normal}.
Interestingly, the critical phase boundary in Fig.~\ref{fig4}(a) turns out to be exponential functions,
which can be strictly proved~\cite{sm}.

Without a magnetic field, the system is
in the $\mathcal{MT}$-symmetric and $\mathcal{MT}$-broken
phases under the $x$-OBC ($\gamma_B=0$) and $x$-PBC ($\gamma_B=1$), respectively;
see Fig.~\ref{fig4}(a). By tuning the
boundary parameter $\gamma_B$,
a continuous $\mathcal{MT}$ transition
connecting two limiting cases can be implemented.
However, varying with $\delta_x$, there is no phase transition
happening in either the $x$-OBC
or the $x$-PBC.
Remarkably, a finite magnetic field
can effectively suppress
the $\mathcal{MT}$-symmetry breaking; see Fig.~\ref{fig4}(b),
which is reflected in two aspects.
First, the $\mathcal{MT}$-symmetric region with large $\gamma_B$
expands with $B$ increased.
Second, the order parameter $d_{\text{HS}}$ diminishes
in the $\mathcal{MT}$-broken region so that
the symmetry breaking becomes weaker, which
is consistent with the magnetic field induced
real Landau levels.
These results reflect the incompatible nature
between the nonreciprocity and the magnetic field.
Such magnetic suppression on the $\mathcal{MT}$-symmetry breaking
indicates that a $\mathcal{MT}$ transition can be
driven by either $\delta_x$ or $B$
for a finite system~\cite{sm}.

In the discussion above, the hopping strength $t_x$ is chosen to be real.
By adding a small imaginary part $i\eta$ to the hopping terms as $\tilde{t}_x^\pm=t+i\eta\pm\delta_x$,
the $\mathcal{MT}$ symmetry of the Hamiltonian in Eq.~\eqref{mt} is destroyed.
As a result, the energy spectra become entirely complex without any $\mathcal{MT}$ transition.
This further proves that the $\mathcal{MT}$ dictates the spectral transition;
see Supplemental Material for details~\cite{sm} .

\emph{Results for nonreciprocal honeycomb lattice.}-
The interplay between the nonreciprocity and a magnetic field
possesses a general picture so that
the physical results are expected to be universal.
To verify this, we perform parallel investigations on the nonreciprocal
honeycomb lattice~\cite{sm}; see Fig.~\ref{fig1}(b),
whose low-energy physics corresponds to the
non-Hermitian massless Dirac particle as sketched in Fig.~\ref{fig1}(d).
We show that the main results obtained in the main text hold true
for the nonreciprocal honeycomb lattice as well~\cite{sm}.
Specifically, the quantization rule persists against
nonreciprocity in the long-wavelength limit,
which gives rise to the familiar Landau levels
$\varepsilon_n^D\propto\pm\sqrt{nB}$ for massless
Dirac particles [cf. Figs.~\ref{fig1}(d)].
The closed cyclotron
orbits formed in the complex space are the physical origin of the
semiclassical quantization.
The $\mathcal{MT}$ phase transition can also be implemented
on the honeycomb lattice, which exhibits similar phase diagrams
and magnetic suppression on the $\mathcal{MT}$-symmetry breaking;
see the Supplemental Material for details~\cite{sm}.

%

\emph{Acknowledgments.}-
We thank Zhong Wang, Chen Fang, Zhesen Yang,
Rui Wang, L. B. Shao, J. L. Lado and Oded Zilberberg
for helpful discussions.
This work was supported by the National
Natural Science Foundation of
China under Grant No. 12074172 (W.C.), No. 12222406 (W.C.) and No. 12174182 (D.Y.X.),
Fundamental Research Funds
for the Central Universities (W.C.), the startup
grant at Nanjing University (W.C.), the State
Key Program for Basic Researches of China
under Grants No. 2021YFA1400403
(D.Y.X.) and the Excellent Programme at Nanjing University.


%

\newpage
\onecolumngrid

\part{\Large{Supplemental Material for ``Cyclotron quantization and mirror-time transition on nonreciprocal lattices''}}

\renewcommand{\thefigure}{S.\arabic{figure}}
\setcounter{figure}{0}
\renewcommand{\theequation}{S.\arabic{equation}}
\setcounter{equation}{0}

\section{I. SUPPRESSION OF NHSE BY A MAGNETIC FIELD}
In this section, we show that a magnetic field
can effectively suppress the NHSE. Its physical reason
is the shrinkage of the point gap for each $k_y$ channel.
We present first the results for the nonreciprocal square lattice
and then those for the nonreciprocal honeycomb lattice.

\subsection{A. Nonreciprocal square lattice}\label{HArper_Lattice}
\begin{figure*}[ht]
\centering
\includegraphics[width=0.7\columnwidth]{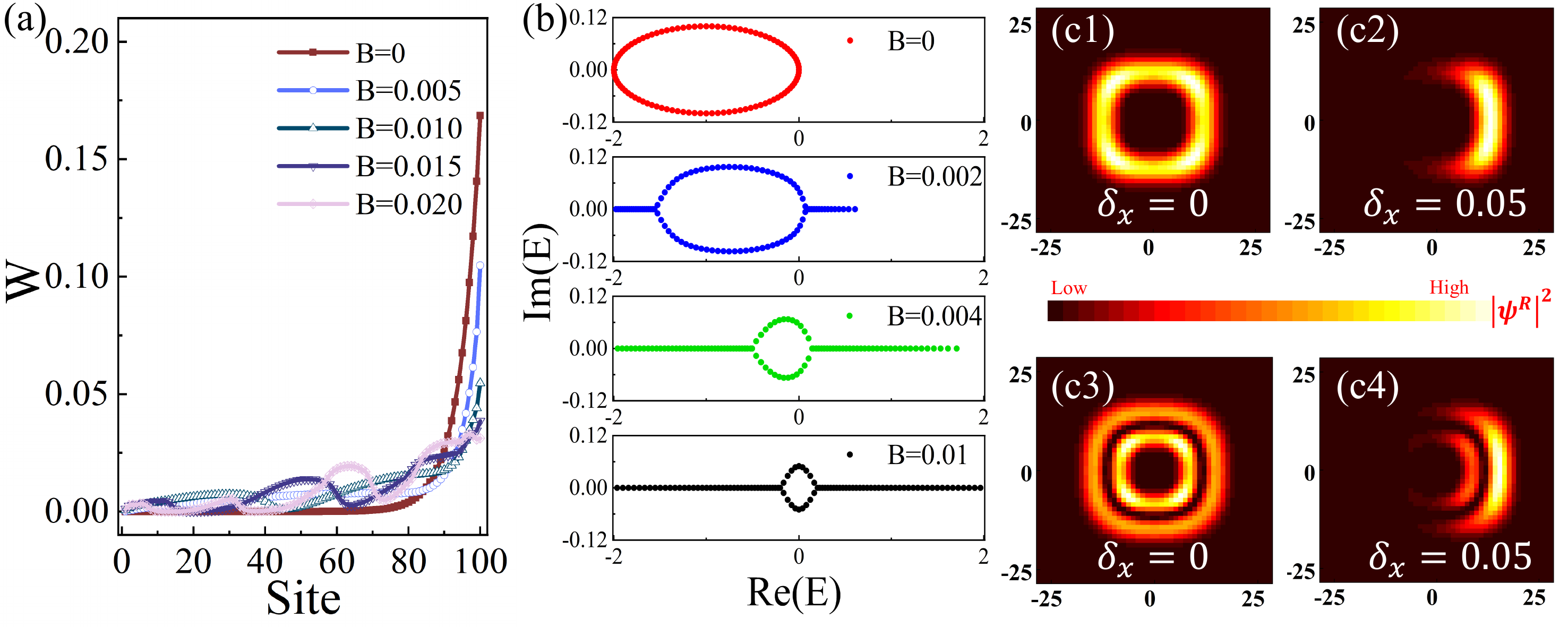}\\
\caption{(a) Spatial distribution functions $W(x,k_y)$ under the $x$-OBC and $y$-PBC.
(b) Complex energy spectra under the $x,y$-PBC for different $B$ with $\delta_x=0.05$, $M=100$ and $k_y=0$.
(c1-c4) Modulation of the wave functions by the nonreciprocal hopping $\delta_x$.
(c1,c2) The 8th and (c3,c4) 58th eigenstates numbered by ascending Re($E$) are randomly chosen
and other parameters are $B=0.02$ and $M=N=50$.
In all figures, $t=0.5$.
}\label{Sfig20}
\end{figure*}

Given that there is no nonreciprocity in the $y$-direction,
meaning that the boundary condition in this direction is unimportant, we
set the $y$-PBC for simplicity. The Hamiltonian can then be
Fourier transformed into $\tilde{H}=-\sum_{m,k_y}\big[t_x^+c^{\dagger}_{m+1,k_y}c_{m,k_y}+t_x^-c^{\dagger}_{m,k_y}c_{m+1,k_y}
+2t\cos {(k_y+2\pi m \phi)}c^{\dagger}_{m,k_y}c_{m,k_y}\big]$.
Taking the $x$-OBC, we plot in Fig.~\ref{Sfig20}(a) the spatial
distribution function $W(x,k_y)=\sum_{i}|\psi^R_i(m,k_y)|^2/M$
defined by all the right eigenstates $\psi^R_i(m,k_y)$
(labeled by $i$) of $\tilde{H}$ for a given $k_y$.
One can see that a small
$B$ is sufficient to drive the
skin modes to penetrate deeply into the bulk,
showing a considerable suppression of the NHSE.
This result generally holds for all transverse
wave vectors $k_y$.

The above results can be understood by the following pictures.
At $B=0$,
the energy spectra $E_{k_y}(k_x)$ for a given $k_y$
forms a closed loop with a point gap topology in its complex
plane under the $x$-PBC; see Fig.~\ref{Sfig20}(b),
indicating the
presence of the
NHSE under the $x$-OBC. For a finite $B$,
real energy spectra develop from the band edges to the center with $B$ increased,
accompanied by a shrinkage of the complex loop; see Fig.~\ref{Sfig20}(b).
According to the correspondence between
the spectra under the $x$-PBC and
the NHSE under the $x$-OBC, this means that
a magnetic field tends to suppress the NHSE.
One can also analyze the results from
the real-space perspective under the $x,y$-OBC
and start with the opposite limit of $\delta_x=0$.
With increasing
$\delta_{x}$ from zero, the wave functions under the magnetic field
are modulated by the exponential envelope function
introduced by the NHSE; see Figs.~\ref{Sfig20}(c1-c4).
Then the results in Fig.~\ref{Sfig20}(a)
can be understood as the superposition of all the broken loops
in real space.

\subsection{B. Nonreciprocal honeycomb lattice}
Next, we investigate the
nonreciprocal honeycomb lattice in Fig.~1(b) of the main text .
With the same Landau gauge $\bm{A}=(0,Bx)$ adopted
and the zigzag edges oriented along the $y$-direction,
the Hamiltonian for the nonreciprocal honeycomb lattice reads
\begin{equation}\label{Graphene_XYH}
\begin{split}
&H'=\sum_{mn}\big(t_x^+b^\dag_{m+1,n}a_{m,n}+t_x^-a^\dag_{m,n}b_{m+1,n}\big)\\
&+t\sum_{mn}\big(e^{i2\pi\phi'm}b^\dag_{m,n}a_{m,n}+e^{-i2\pi\phi'm}b^\dag_{m,n+1}a_{m,n}+\text{H.c.}\big),
\end{split}
\end{equation}
where $a^{\dagger}_{m,n}, b^\dag_{m,n}$ ($a_{m,n}, b_{m,n}$) are the creation (annihilation)
operators for the A, B sublattices, respectively,
and $\bm{R}_{(m,n)}=m\bm{a}_1+n\bm{a}_2$ is the location of
the lattice sites with $(\bm{a}_{1},\bm{a}_{2})$ the unit
vectors shown in Fig.~1(b) of the main text.
The phase factor is defined by $\phi'=\Phi'/(2\Phi_0)$
with $\Phi'=3\sqrt{3}Ba'^2/2$ the flux through a unit cell
and $a'$ the bond length that is set to $a'=1$ henceforth.

\begin{figure}[ht]
\centering
\includegraphics[width=0.7\columnwidth]{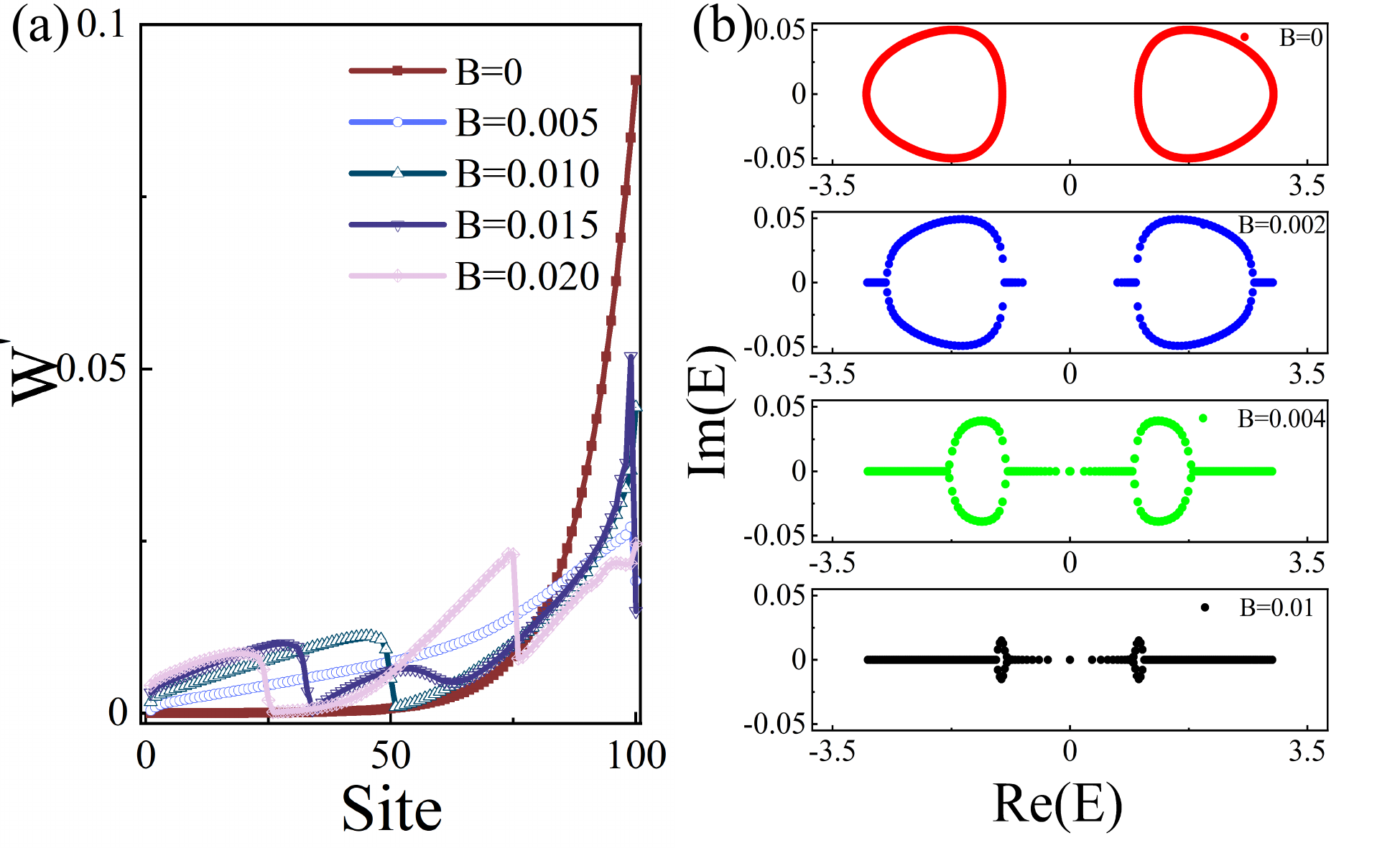}\\
\caption{(a) The spatial distribution functions $W'(x,k_y)$
under the $x$-OBC and $y$-PBC
and (b) the complex energy spectra under the $x$-PBC and $y$-PBC for different
$B$ with $\delta_x=0.05$, $t=1$, $M=200$ and $k_y=0$.}\label{fig_Graphene_PO_Point}
\end{figure}

Similar to the square lattice, we take the $y$-PBC
and rewrite the Hamiltonian into
$
\tilde{H}'=\sum_{m,k_y}\big[ \Delta a_{m,k_y}^{\dagger}b_{m,k_y}+\Delta^* b_{m,k_y}^{\dagger}a_{m,k_y}+t_x^+b_{m+1,k_y}^{\dagger}a_{m,k_y}
+t_x^-a_{m,k_y}^{\dagger}b_{m+1,k_y}\big]
$ with $\Delta=2t\cos{[\sqrt{3}k_y/2+\pi\phi(m-5/6)]}$.
The spatial distribution function is calculated by  $W'(x,k_y)=\sum_{i}
(|\psi_{a,i}^{R}(m,k_y)|^2+|\psi_{b,i}^{R}(m,k_y)|^2)/(2M)$ with
$\psi_{a,i}^{R}(m,k_y)$ and $\psi_{b,i}^{R}(m,k_y)$ the
components on the A and B sublattices, respectively.
The spatial distribution $W'(x,k_y)$ is plotted
in Fig.~\ref{fig_Graphene_PO_Point}(a). Similar to the results of the square lattice,
the skin modes are strongly suppressed by just a small $B$.
It is closely related to the shrinkage of two complex loops
of the energy spectra under the $x$-PBC (for an arbitrary $k_y$);
see Fig.~\ref{fig_Graphene_PO_Point}(b). The conclusions
agree with those of the square lattice in Sec.I.A.

\section{II. SCALING OF LOW-ENERGY LANDAU LEVELS}\label{B}

\begin{figure}[h!]
\centering
\includegraphics[width=\columnwidth]{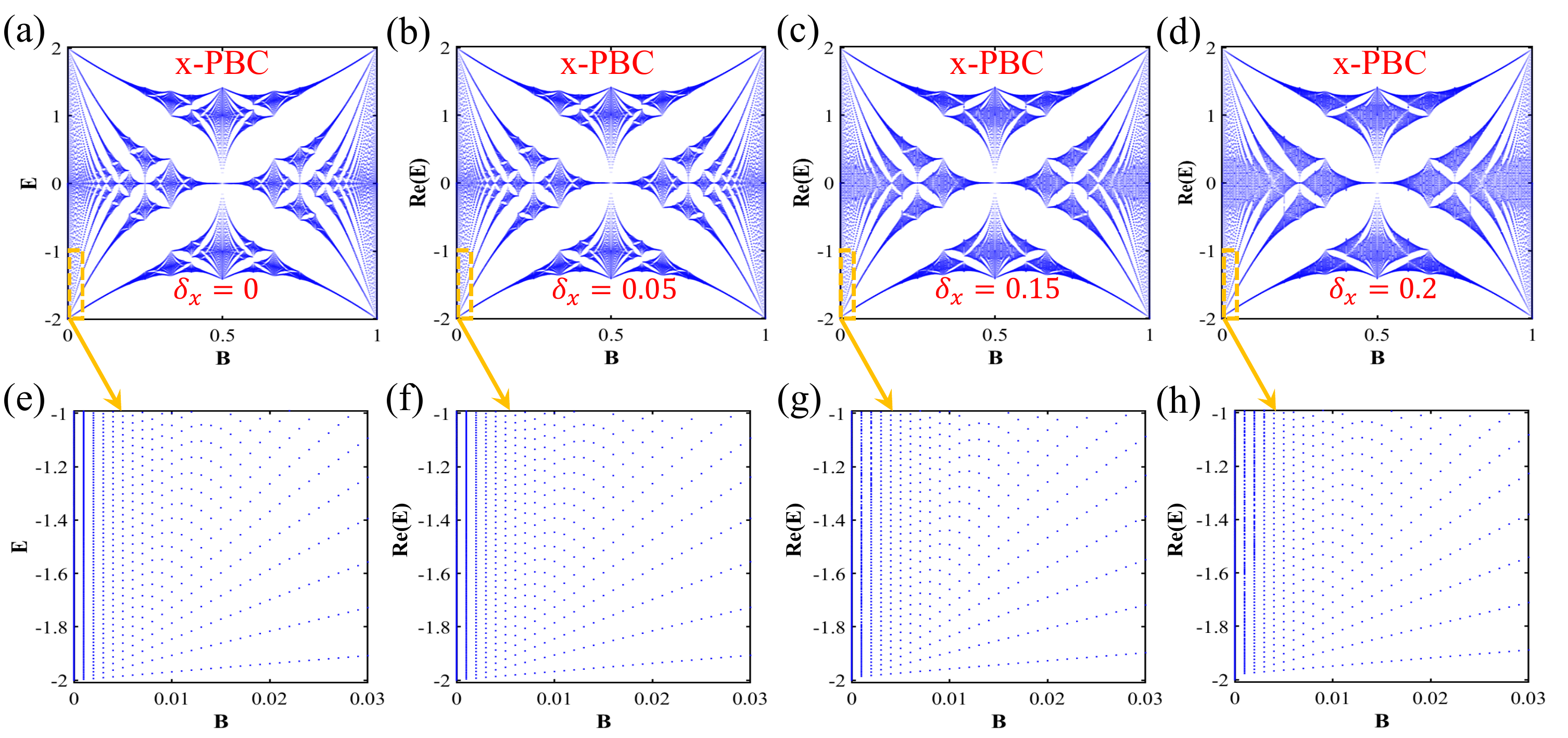}\\
\caption{(a-d) Energy spectra under the $x$-PBC and $y$-PBC with $\delta_x=0,
0.05, 0.15, 0.2$ and $t=0.5$. (e-h)
Zoom of the Landau fan structures corresponding to (a-d).   }\label{figs2}
\end{figure}

Fig.~\ref{figs2} shows the energy
spectra for different
$\delta_x$ under the $x$-PBC.
One can see that the spectra
undergo a visible modification as $\delta_x$ increases.
Interestingly, the Landau fan structures with
real energy values persist
in the long-wavelength limit.
Figs.~\ref{figs2}(e-h) show that the Landau
levels exhibit the scaling $E_n\propto nB$,
which resembles the normal particle behavior.
Such numerical results can be well explained
by the semiclassical quantization introduced in
the next section.

\section{III. SEMICLASSICAL ORBITS AND QUANTIZATION CONDITIONS}
In this section, we derive the time-dependent coordinate functions and the trajectories of the
wave packet under a magnetic field based on the semiclassical equation of motion for both
the nonreciprocal square and honeycomb lattices. We show that, in the long-wavelength limit,
the semiclassical orbits are always closed loops, which give rise to real Landau levels.
Beyond the long-wavelength limit, the orbits become open and the energies are complex.

\subsection{A. Nonreciprocal square lattice}
Without a magnetic field, the eigenenergy of Hamiltonian~(1) in the main text is
\begin{equation}\label{Dispersion_B0}
\begin{split}
E(\bm{k})=-2t(\cos k_x+\cos k_y)-2i\delta_x\sin k_x
\end{split}
\end{equation}
The Hamiltonian under a magnetic field can be obtained
through the Peierls substitution ($\hbar k\rightarrow \bm{p}-q\bm{A}$) as
\begin{equation}\label{Hamiltonian_Peierls}
\begin{split}
h=-2\left[t\cos p_x+t\cos(p_y-Bx)+i\delta_x\sin{p_x}\right],
\end{split}
\end{equation}
where $\bm{A}=(0,Bx)$ and $\hbar=q=a=1$ have been adopted. The Hamilton's canonical equations read
\begin{equation}\label{Hamil_Patrial_EQ0}
\begin{split}
\frac{dx}{d\tau}&=\frac{\partial h}{\partial p_x}=2t(\sin{p_x}-i\delta_x \cos{p_x}),\\
\frac{dy}{d\tau}&=\frac{\partial h}{\partial p_y}=2t\sin{(p_y-Bx)},\\
\frac{dp_x}{d\tau}&=-\frac{\partial h}{\partial x}=2Bt\sin{(p_y-Bx)},\\
\frac{dp_y}{d\tau}&=-\frac{\partial h}{\partial y}=0.\\
\end{split}
\end{equation}
One can obtain the coordinate functions and trajectories of the wave packet with given initial conditions of
the coordinates and canonical momenta. In general, the above differential equations
can only be solved numerically.

We are mainly interested in the long-wavelength limit,
where expanding $E(\bm{k})$ around $(k_x,k_y)=(0,0)$
yields $\varepsilon(\bm{k})= t(k_x^2+k_y^2)+2i\delta_x k_x$.
Accordingly, Eq.~\eqref{Hamil_Patrial_EQ0} reduce to

\begin{equation}\label{Hamil_Patrial_EQ1}
\begin{split}
\frac{dx}{d\tau}&=2(tp_x-i\delta_x),\\
\frac{dy}{d\tau}&=2t(p_y-Bx),\\
\frac{dp_x}{d\tau}&=2Bt(p_y-Bx),\\
\frac{dp_y}{d\tau}&=0,\\
\end{split}
\end{equation}
which can be solved analytically.
By eliminating the momenta in Eq.~\eqref{Hamil_Patrial_EQ1} we obtain the differential equations
solely for the coordinates as
\begin{equation}\label{traject_Patrial_EQ}
\begin{split}
\frac{d^2x}{d\tau^2}&=2Bt\frac{dy}{d\tau},\\
\frac{d^2y}{d\tau^2}&=-2Bt\frac{dx}{d\tau}.\\
\end{split}
\end{equation}
The solutions of Eq.~\eqref{traject_Patrial_EQ} have the form of
\begin{equation}\label{Trail_Solutions}
\begin{split}
x(\tau)&=A_1e^{i\omega \tau}+A_2e^{-i\omega \tau}+x_0,\\
y(\tau)&=i(A_1e^{i\omega \tau}-A_2e^{-i\omega \tau})+y_0,
\end{split}
\end{equation}
where $\omega=2Btq/\hbar^2$ is the cyclotron frequency and the
four parameters are determined by the initial conditions for
the coordinates $x(0),y(0)$ and those for the momenta $p_x(0),p_y(0)$ through
\begin{equation}
\begin{split}
A_1&=\frac{t\left[Bx(0)-ip_x(0)-p_y(0)\right]-\delta_x}{2Bt},\\
A_2&=\frac{t\left[Bx(0)+ip_x(0)-p_y(0)\right]+\delta_x}{2Bt},\\
x_0&=p_y(0)/B,\ \
y_0=y(0)-\frac{tp_x(0)-i\delta_x}{Bt}.
\end{split}
\end{equation}
The periodic functions in Eq.~\eqref{Trail_Solutions} imply
that a wave packet always forms closed orbits in the complex $x$-$y$ space.
Specifically, the projections of the trajectories in the
$\text{Re}(x)$-$\text{Re}(y)$ and $\text{Im}(x)$-$\text{Im}(y)$
planes are closed loops described by the equations as
\begin{equation}\label{Traject_Equation_Circle}
\begin{split}
&\text{Re}:\left[x-\frac{p_y(0)}{B}\right]^2+\left[y-y(0)+
\frac{p_x(0)}{B}\right]^2=\left(\frac{A}{B}\right)^2,\\
&\text{Im}:x^2+\left(y-\frac{\delta_x}{Bt}\right)^2=\left(\frac{\delta_x}{Bt}\right)^2,\\
\end{split}
\end{equation}
where $A=\sqrt{[Bx(0)-p_{y}(0)]^2+p_{x}(0)^2}$.
The closed orbits of the wave packet under a magnetic field
mean that the quantization rule must be maintained due to the single-valued nature of the wave functions,
which determines the energy values. Here, it is just the
Onsager-Lifshitz quantization rule
\begin{equation}\label{Onsager}
\begin{split}
\oint \bm{p}\cdot d\bm{r}= (n+\frac{1}{2})h.
\end{split}
\end{equation}
From Eq.~\eqref{Hamil_Patrial_EQ1}, the relations between canonical momenta and velocities are
\begin{equation}\label{PXPY}
\begin{split}
p_x&=\frac{1}{2t}\frac{dx}{d\tau}+\frac{i\delta_x}{t},\\
p_y&=\frac{1}{2t}\frac{dy}{d\tau}+Bx.\\
\end{split}
\end{equation}
By inserting Eq.~\eqref{PXPY} into $\varepsilon(\bm{p}-q\bm{A})$ we obtain the energy as
\begin{equation}\label{Energy_CA}
\begin{split}
\varepsilon=\frac{1}{4t}\left[(\frac{dx}{d\tau})^2+
(\frac{dy}{d\tau})^2\right]+\frac{\delta_x^2}{t}
=\frac{\omega^2}{t}A_1A_2+\frac{\delta_x^2}{t}.
\end{split}
\end{equation}
Meanwhile, inserting Eq.~\eqref{PXPY} into the integral of Eq.~\eqref{Onsager} yields
\begin{equation}\label{Onsagers}
\begin{split}
\oint \bm{p}\cdot d\bm{r}=\oint\bm{p}\cdot \frac{d\bm{r}}{d\tau}d\tau
=\frac{2\pi\omega}{t}A_1A_2= (n+\frac{1}{2})h.
\end{split}
\end{equation}
By combining Eqs.~\eqref{Energy_CA} and ~\eqref{Onsagers}
we finally obtain the real Landau levels as
\begin{equation}\label{Square_Landau_Level}
\begin{split}
\varepsilon_n=(n+\frac{1}{2})\hbar\omega+\frac{\delta_x^2}{t},
\end{split}
\end{equation}
which deviates from the standard results of normal particles by a factor $\delta_x^2/t$ stemming
from the nonreciprocity. We conclude that, in the long wave-length limit,
the closed orbits of the wave packet impose the quantization rule,
which preserves real Landau levels despite the nonreciprocity.
In contrast, semiclassical orbits solved numerically by Eq.~\eqref{Hamil_Patrial_EQ0} are open lines
beyond the long-wavelength limit, which are shown in Fig.~3(b) of the main text. As a result,
the quantization rules break down and the spectra become complex.

\subsection{B. Nonreciprocal honeycomb lattice}
\begin{figure}[h]
\centering
\includegraphics[width=0.7\columnwidth]{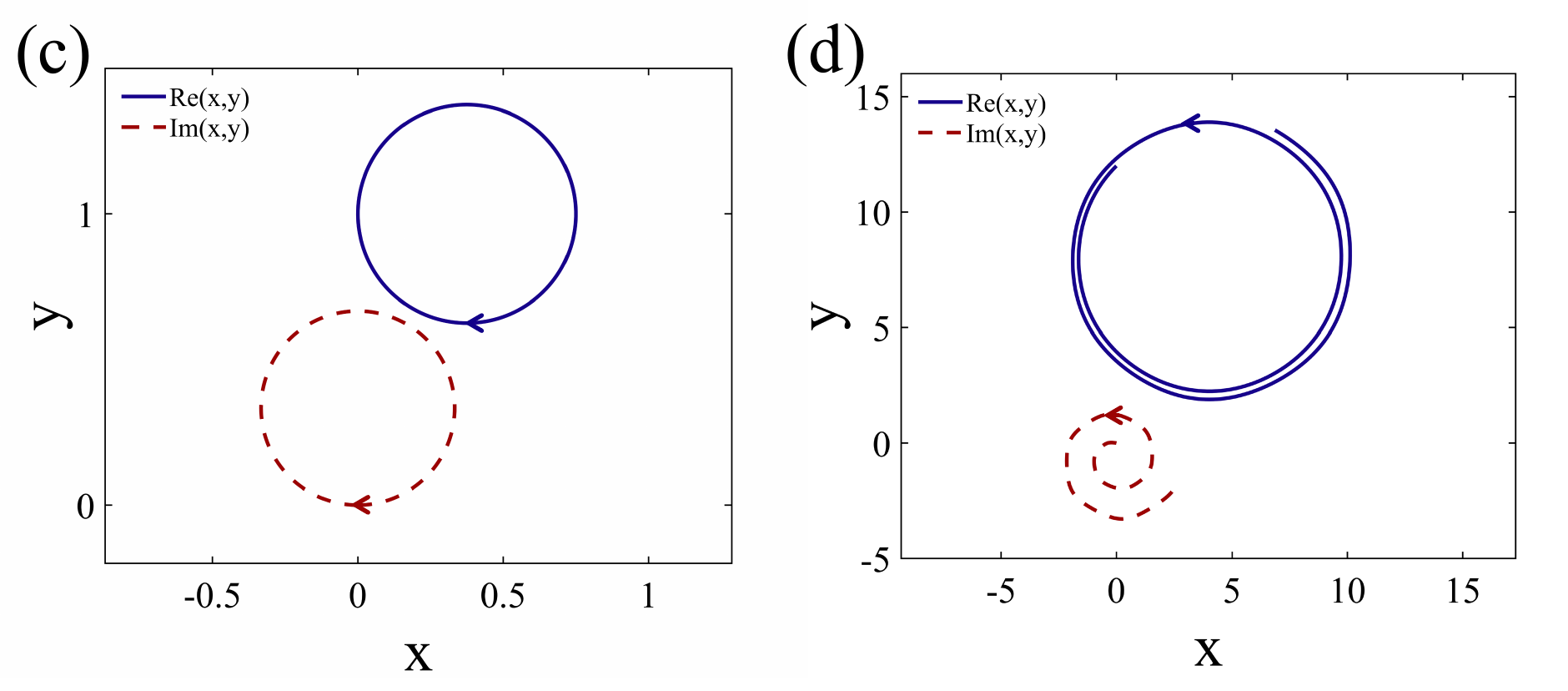}\\
\caption{Projections of the semiclassical trajectories in the real
and imaginary $x$-$y$ planes for the nonreciprocal honeycomb lattice. (a) Closed orbits in the
long-wavelength limit with $B=0.4$, $x(0)=0$,
$y(0)=1$, $p_{x}(0)=0$, $p_{y}(0)=-0.15$, and (b) open trajectories beyond this
limit with $B=0.2$, $x(0)=0$, $y(0)=12$, $p_{x}(0)=0.8$, $p_{y}(0)=0.8$.  Other parameter is $\delta_x=0.2$.  }
\label{LOOPS_Graphene}
\end{figure}

In this subsection, we derive the semiclassical
trajectories of the wave packet on the nonreciprocal honeycomb lattice.
Without a magnetic field, the Bloch Hamiltonian reads
\begin{equation}\label{Bloch_Graphene}
\begin{split}
H'(\bm{k})&=\sum_{k_x,k_y}\left(
    \begin{array}{cc}
      0 & t_x^+e^{-i\bm{k}\cdot\bm{\Delta}_1}+te^{-i\bm{k}\cdot\bm{\Delta}_2}+te^{-i\bm{k}\cdot\bm{\Delta}_3} \\
      t_x^-e^{i\bm{k}\cdot\bm{\Delta}_1}+te^{i\bm{k}\cdot\bm{\Delta}_2}+te^{i\bm{k}\cdot\bm{\Delta}_3} & 0 \\
    \end{array}
  \right)
  c_{k_x,k_y}^{\dagger}c_{k_x,k_y},\\
\end{split}
\end{equation}
where $\bm{\Delta}_1=(1,0),\bm{\Delta}_2=(-\frac{1}{2},\frac{\sqrt{3}}{2}),
\bm{\Delta}_3=(-\frac{1}{2},-\frac{\sqrt{3}}{2})$. The conduction and valence bands
possess the following dispersion
\begin{equation}\label{Dispersion_Graphene}
\begin{split}
E'_\pm(\bm{k})=\pm\sqrt{t^2-\delta_x^2+4t^2\cos^2{\frac{\sqrt{3}k_y}{2}}+
4t^2\cos{\frac{3k_y}{2}}\left(\cos{\frac{3k_x}{2}
-\frac{\delta_x}{t}\sin{\frac{3k_x}{2}}}\right)}.
\end{split}
\end{equation}
We expand the expressions around two Dirac points $\mathbf{K}=(0,\frac{4\pi}{3\sqrt{3}}), \mathbf{K}'=(0,-\frac{4\pi}{3\sqrt{3}})$
to study the physics in the long-wavelength limit. The energies reduce to
\begin{equation}\label{Dispersion_Graphene_Expand}
\begin{split}
\varepsilon'_\pm(\bm{k})=\pm\frac{3t}{2}\sqrt{k_x^2+k_y'^{2}-\left(\frac{2\delta_x}{3t}\right)^2+
\frac{4i\delta_x}{3t}k_x},
\end{split}
\end{equation}
where $k_y'=k_y-\frac{4\pi}{3\sqrt{3}}$ is measured from the Dirac points. For $\delta_x=0$, Eq.~\eqref{Dispersion_Graphene_Expand}
describes massless Dirac particles. The semiclassical Hamiltonian for the conduction
band under a magnetic field is modified into
\begin{equation}\label{Graphene_Sem_Ham2}
\begin{split}
h'=t_0\sqrt{\left(p_x+i\frac{\delta_x}{t_0}\right)^2+(p_y-Bx)^2}=t_0h_0,
\end{split}
\end{equation}
with $t_0=3t/2$ and $h_0=\sqrt{(p_x+i\frac{\delta_x}{t_0})^2+(p_y-Bx)^2}$.
The Hamilton's canonical equations read
\begin{equation}\label{Hamil_Gra_Patrial_EQ2}
\begin{split}
\frac{dx}{d\tau}&=t_0\frac{(p_x+i\frac{\delta_x}{t_0})}{h_0},\\
\frac{dy}{d\tau}&=t_0\frac{(p_y-Bx)}{h_0},\\
\frac{dp_x}{d\tau}&=t_0\frac{B(p_y-Bx)}{h_0},\\
\frac{dp_y}{d\tau}&=0.\\
\end{split}
\end{equation}
Eliminating the momenta in Eq.~\eqref{Hamil_Gra_Patrial_EQ2} we obtain the differential equations solely for the coordinates as
\begin{equation}\label{Hamil_Patrial_EQ3}
\begin{split}
\frac{d^2x}{d\tau^2}&=B\frac{t_0}{h_0}\frac{dy}{d\tau},\\
\frac{d^2y}{d\tau^2}&=-B\frac{t_0}{h_0}\frac{dx}{d\tau}.\\
\end{split}
\end{equation}
The solutions of Eq.~\eqref{Hamil_Patrial_EQ3} have the form of
\begin{equation}\label{Trail_Solutions2}
\begin{split}
x(\tau)&=A_1'e^{i\omega' \tau}+A_2'e^{-i\omega' \tau}+x_0',\\
y(\tau)&=i(A_1'e^{i\omega' \tau}-A_2'e^{-i\omega' \tau})+y_0',
\end{split}
\end{equation}
where $\omega'=qBt_0^2/(\hbar^2h_0)$ is the cyclotron frequency and the
four parameters are determined by the initial conditions through
\begin{equation}\label{Factor_Graphene_trajec}
\begin{split}
A_1'&=\frac{t_0[Bx(0)-ip_x(0)-p_y(0)]+\delta_x}{2Bt_0},\\
A_2'&=\frac{t_0[Bx(0)+ip_x(0)-p_y(0)]-\delta_x}{2Bt_0},\\
x_0'&=p_y(0)/B,\ \
y_0'=y(0)-\frac{t_0p_x(0)+i\delta_x}{Bt_0}.
\end{split}
\end{equation}
The coordinate functions possess the same form as
those of the normal particles in Eq.~\eqref{Trail_Solutions} so that the trajectories
form closed orbits as well.
The projections of the trajectories in the
$\text{Re}(x)$-$\text{Re}(y)$ and $\text{Im}(x)$-$\text{Im}(y)$
planes are closed loops described by the equations as
\begin{equation}\label{Traject_Equation_Circle}
\begin{split}
&\text{Re}:\left[x-\frac{p_y(0)}{B}\right]^2+\left[y-y(0)+
\frac{p_x(0)}{B}\right]^2=\left(\frac{A}{B}\right)^2,\\
&\text{Im}:x^2+\left(y+\frac{\delta_x}{Bt_0}\right)^2=
\left(\frac{\delta_x}{Bt_0}\right)^2,\\
\end{split}
\end{equation}
with $A=\sqrt{[Bx(0)-p_{y}(0)]^2+p_{x}(0)^2}$; see Fig.~\ref{LOOPS_Graphene}(a).
The same results hold true for the valence band as well.
Similarly, closed orbits impose the following
Onsager-Lifshitz quantization rule as
\begin{equation}\label{Onsager2}
\begin{split}
\oint \bm{p}\cdot d\bm{r}=(n+\frac{1}{2}-\gamma)h,
\end{split}
\end{equation}
in which the factor $\gamma=\frac{1}{2}$ is due to the Berry phase
of the Dirac particles, different from the case of normal particles.

We will show that the quantization also gives
rise to real Landau levels.
From Eq.~\eqref{Hamil_Gra_Patrial_EQ2}, we have
\begin{equation}\label{PXPY2}
\begin{split}
p_x&=\frac{h_0}{t_0}\frac{dx}{d\tau}-\frac{i\delta_x}{t_0},\\
p_y&=\frac{h_0}{t_0}\frac{dy}{d\tau}+Bx.\\
\end{split}
\end{equation}
By inserting Eq.~\eqref{PXPY2} into Eq.~\eqref{Graphene_Sem_Ham2}, the energy can be expressed as
\begin{equation}\label{Energy_Gra_CA}
\begin{split}
\varepsilon'_\pm=\pm h_0\sqrt{(\frac{dx}{d\tau})^2+(\frac{dy}{d\tau})^2}
=\pm2h_0\sqrt{A_1A_2}\omega.\\
\end{split}
\end{equation}
Meanwhile, inserting Eq.~\eqref{PXPY2} into the integral in Eq.~\eqref{Onsager2} yields
\begin{equation}\label{diracquanti}
\begin{split}
\oint \bm{p}\cdot d\bm{r}=\oint\bm{p}\cdot \frac{d\bm{r}}{d\tau}d\tau
=\frac{2\pi\omega h_0}{t_0}4A_1A_2=nh.
\end{split}
\end{equation}
By combining Eqs.~\eqref{Energy_Gra_CA} and ~\eqref{diracquanti}
we obtain the real Landau levels as
\begin{equation}\label{Square_Landau_Level}
\begin{split}
\varepsilon_n^D=\varepsilon'_{\pm n}=\pm\frac{3}{2}t\sqrt{q\hbar n B},
\end{split}
\end{equation}
which are just the familiar Landau levels for massless Dirac
particles.

For more general cases, the Hamilton's canonical equations read
\begin{equation}\label{Hamil_Patrial_EQ}
\begin{split}
\frac{dx}{d\tau}&=\frac{3it\cos{[\frac{\sqrt{3}(p_y-Bx)}{2}]}
e^{-\frac{3ip_x}{2}}(-t+e^{3ip_x}(t-\delta_x)-\delta_x)}{2E'_+(\bm{p}-q\bm{A})},\\
\frac{dy}{d\tau}&=-\frac{\sqrt{3}te^{-\frac{3ip_x}{2}}\sin{[\frac{\sqrt{3}(p_y-Bx)}{2}]}
(t+e^{3ip_x}(t-\delta_x)+\delta_x+
4te^{\frac{3ip_x}{2}}\cos{[\frac{\sqrt{3}(p_y-Bx)}{2}]})}
{2E'_+(\bm{p}-q\bm{A})},\\
\frac{dp_x}{d\tau}&=-\frac{\sqrt{3}Bte^{-\frac{3ip_x}{2}}\sin{[\frac{\sqrt{3}(p_y-Bx)}{2}]}
(t+e^{3ip_x}(t-\delta_x)+\delta_x+
4te^{\frac{3ip_x}{2}}\cos{[\frac{\sqrt{3}(p_y-Bx)}{2}]})}
{2E'_+(\bm{p}-q\bm{A})},\\
\frac{dp_y}{d\tau}&=0,
\end{split}
\end{equation}
with $E'_+(\bm{p}-q\bm{A})=\sqrt{t^2-\delta_x^2+4t^2\cos^2{[\frac{\sqrt{3}(p_y-Bx)}{2}]}+
4t^2\cos{[\frac{3(p_y-Bx)}{2}]}(\cos{\frac{3p_x}{2}
-\frac{\delta_x}{t}\sin{\frac{3p_x}{2}}})}$.
We solve these equations numerically and plot the semiclassical trajectories that are open lines
in Fig.~\ref{LOOPS_Graphene}(b). Therefore,
the quantization conditions disappear and the energy spectra become
complex beyond the long-wavelength limit.

\section{IV. ENERGY SPECTRA OF NONRECIPROCAL SQUARE LATTICE UNDER $x$-OBC}\label{A}
\begin{figure}[h]
\centering
\includegraphics[width=\columnwidth]{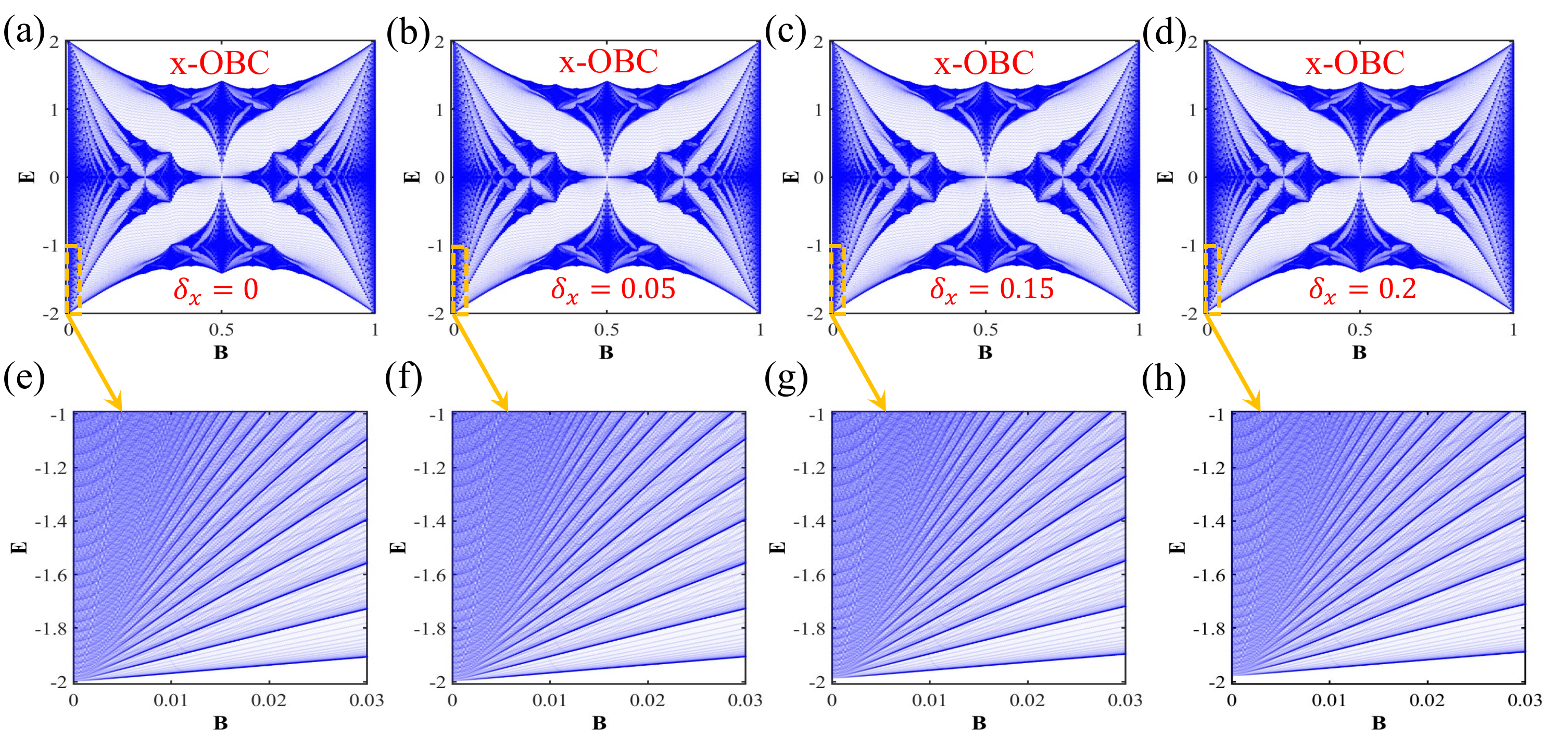}\\
\caption{(a-d) Energy spectra under the
$x$-OBC and $y$-PBC for $\delta_x=0, 0.05, 0.15, 0.2$ with $M=50$, $t=0.5$.
(e-h) Zoom of the Landau fan structures corresponding to (a-d).   }
\label{Figs1}
\end{figure}

Fig.~\ref{Figs1} shows the energy
spectra for different strengths of the nonreciprocal hopping
$\delta_x\in[0,2t/5]$ under the $x$-OBC, where the in-gap streaks
are edge states.
The energy spectra are entirely real and
exhibit a weak dependence on $\delta_x$.
Moreover, the Landau fan in the long-wavelength
limit exhibits equal level spacing
and linear dependence on $B$,
which reduces to the behavior
of free particles with a quadratic
dispersion, the same as that under the $x$-PBC.

\section{V. PROOF OF THE EXPONENTIAL PHASE BOUNDARY}
We have seen in Fig.~4 of the main text that a $\mathcal{MT}$
transition can be induced by both the boundary parameter
$\gamma_B$ and the magnetic field $B$ for systems with a finite size
in the $x$-direction.
The critical phase boundary can always be well fitted
by exponential functions. Here, we prove this result for $B=0$.
In this case, the Fourier transformed ($y$-direction) Hamiltonian reduces to
\begin{equation}\label{Hamil_Mky}
\begin{split}
\tilde{H}=t_x^+c^{\dagger}_{m+1}c_{m}+t_x^-c^{\dagger}_{m}c_{m+1}
+2tc^{\dagger}_{m}c_{m}+\gamma_B(t_x^+c^{\dagger}_{1}c_{M}+t_x^-c^{\dagger}_{M}c_{1}),
\end{split}
\end{equation}
where the original $k_y$ dependent term that is irrelevant
to the phase boundary has been dropped and the tunable boundary hopping is introduced.
\par The eigenvalue equation $\tilde{H}|\Psi\rangle=E|\Psi\rangle$ with $|\Psi\rangle=\sum_m\psi_m|m\rangle$ and $|m\rangle=c^{\dagger}_{m}|0\rangle \ \ (m=1,\ldots,M)$
consists of the bulk equations
\begin{equation}\label{Bulk_eq}
\begin{split}
t_x^+\Psi_{j-1}-E\Psi_{j}+t_x^-\Psi_{j+1}=0
\end{split}
\end{equation}
with $j=2,\ldots,M-1$, and the boundary equations
\begin{equation}
\begin{split}
-E\Psi_{1}+t_x^-\Psi_{2}+\gamma_Bt_x^+\Psi_{M}=0, \\
\gamma_Bt_x^-\Psi_{1}+t_x^+\Psi_{M-1}-E\Psi_{M}=0.
\end{split}
\end{equation}
Due to the spatially translational invariance of bulk equations, we can take the ansatz of wave function $\Psi_{i}$ as:
\begin{equation}\label{ans_Wave}
\begin{split}
\Psi_{i}=(\beta_i,\beta_i^2,\beta_i^3,\cdots,\beta_i^{M-1},\beta_i^{M})^T.
\end{split}
\end{equation}
From Eq.~\eqref{Bulk_eq} and Eq.~\eqref{ans_Wave}, we obtain the eigenvalue in terms of $\beta_i$ as
\begin{equation}\label{Eigenvalue1}
\begin{split}
E=\frac{t_x^+}{\beta_i}+t_x^-\beta_i.
\end{split}
\end{equation}
For any $E$, there are two solutions $\beta_i=\beta_1, \beta_2$ that fulfill the constraint
\begin{equation}\label{Eigenvalue2}
\begin{split}
\beta_1\beta_2=\frac{t_x^+}{t_x^-}.
\end{split}
\end{equation}
Note that any superposition of the two linearly independent solutions $\Psi=b_1\Psi_1+b_2\Psi_2=(\psi_1,\psi_2,\cdots,\psi_M)$ is also a solution of Eq.~\eqref{Bulk_eq}, where $\psi_m=b_1\beta_1^m+b_2\beta_2^m$.

Inserting $\Psi$ into the boundary equations yields
\begin{equation}\label{Boundary_Psi}
\begin{split}
(\beta_1^{M+1}+\beta_2^{M+1})-(\frac{t_x^+\gamma_B^2}{t_x^-})(\beta_1^{M-1}-\beta_2^{M-1})
-\Big[(1+\Big(\frac{t_x^+}{t_x^-}\Big)^M\Big]\gamma_B(\beta_1-\beta_2)=0.
\end{split}
\end{equation}
It is convenient to set the two solutions as
\begin{equation}\label{Zr}
\begin{split}
\beta_1=re^{i\theta},\ \ \beta_2=re^{-i\theta}
\end{split}
\end{equation}
with $r=\sqrt{t_x^+/t_x^-}$, which fulfills Eq.~\ref{Eigenvalue2}. Then Eq.~\eqref{Boundary_Psi} reduces  to
\begin{equation}\label{eq_SIN}
\begin{split}
\sin[(M+1)\theta]-\eta_1\sin[(M-1)\theta]-\eta_2\sin\theta=0,
\end{split}
\end{equation}
with $\eta_1=\gamma_B^2$ and $\eta_2=\gamma_B(r^{-M}+r^{M})$,
and the eigenvalues become
\begin{equation}\label{E_theta}
\begin{split}
E=2\sqrt{t_x^+t_x^-}\cos\theta.
\end{split}
\end{equation}
The eigenvalue $E$ may be real or complex
depending on the solutions of $\theta$ in Eq.~\eqref{eq_SIN}.

For $\gamma_B=0$, i.e., the $x$-OBC, we have $\eta_1=\eta_2=0$, Eq.~\eqref{eq_SIN} reduces to
\begin{equation}\label{eq_SIN_GB0}
\begin{split}
\sin[(M+1)\theta]=0,
\end{split}
\end{equation}
which yields $M$ real roots as $\theta=l\pi/(M+1)$ with $l=1,\ldots,M$.
As a result, the system is in the $\mathcal{MT}$-symmetric phase
with entirely real energy spectra under the $x$-OBC.

For $\gamma_B\neq0$, we rewrite Eq.~\eqref{eq_SIN} as $F_1(\theta)=F_2(\theta)$
with
\begin{equation}\label{F1F2}
\begin{split}
F_1(\theta)=\sin[(M+1)\theta]-\eta_1\sin[(M-1)\theta],\ \
F_2(\theta)=\eta_2\sin\theta.
\end{split}
\end{equation}
The eigenvalues $E$ are determined by the solutions of $F_1(\theta)=F_2(\theta)$,
which correspond to the crossing points of the two functions. It can be shown that
as long as $\eta_2<1+\eta_1$, there exist $M$ real solutions for $\theta$
and $E$; Otherwise, as $\eta_2>1+\eta_1$, some of the solutions of
$\theta$ and $E$ become complex~\cite{Guo21prl}. As a result,
$\eta_2=1+\eta_1$ gives the $\mathcal{MT}$ transition point,
which determines the critical value $\gamma_B^c=r^{-M}$ for $\gamma_B\leq1$.
The phase boundary possesses the asymptotic form for $\delta_x\ll t$ as
\begin{equation}\label{Exp_gammaB}
\begin{split}
\gamma^c_B=e^{-\delta_x M/t},
\end{split}
\end{equation}
showing that it is an exponential function.

\section{VI. SIZE EFFECT OF THE $\mathcal{MT}$ TRANSITION}
\begin{figure}[h]
\centering
\includegraphics[width=0.7\textwidth]{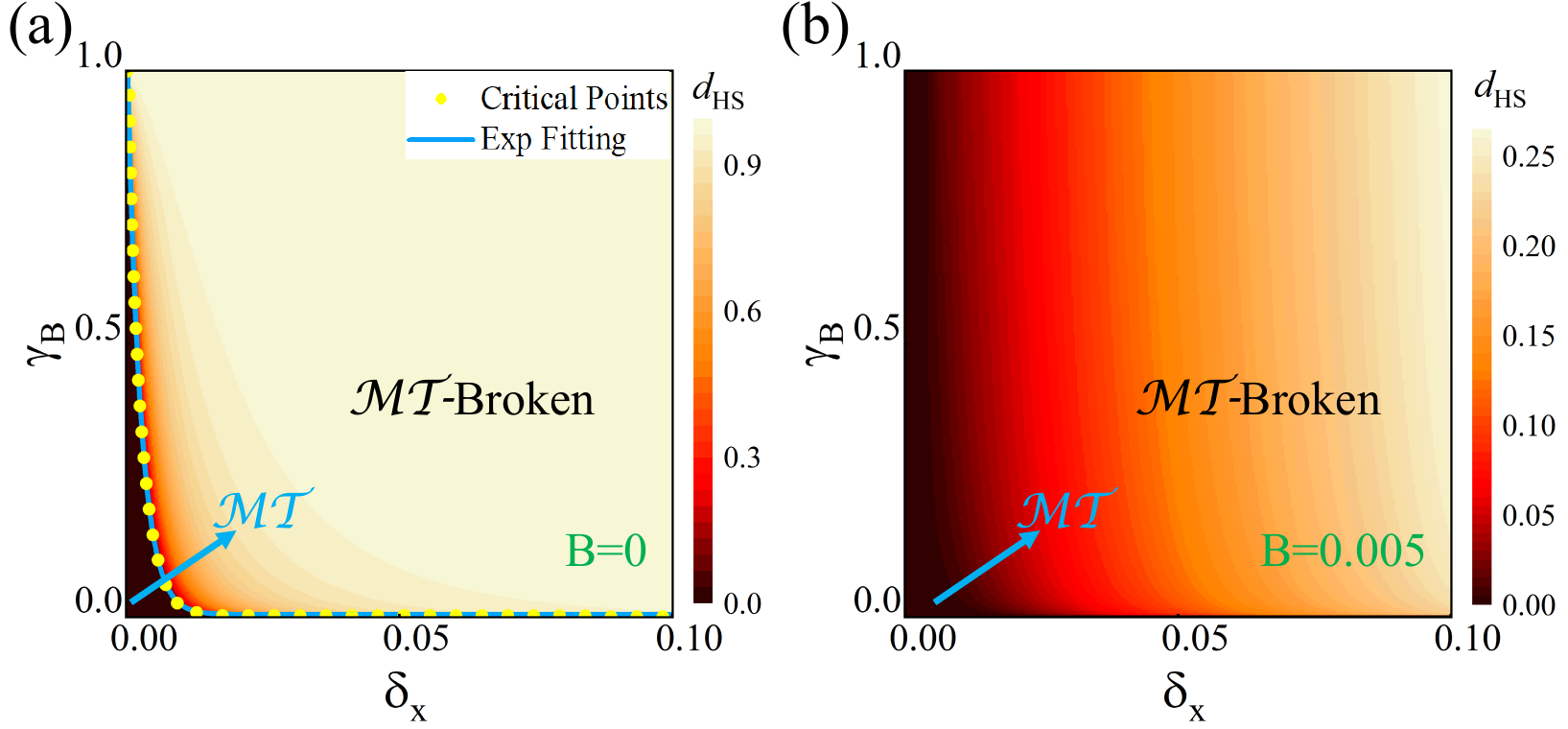}\\
\caption{$\mathcal{MT}$ phase diagram for $M=200$.
Other parameters are the same as those in Fig.~5 of the main text.}\label{Sfig3}
\end{figure}

Eq.~\eqref{Exp_gammaB} also tells that the function of the phase boundary $\gamma_B^c(\delta_x)$
strongly depends on the system size $M$ in the $x$-direction.
Specifically, the area of the $\mathcal{MT}$-symmetric phase
reduces as $M$ increases.
In Fig.~\ref{Sfig3}, we plot the phase diagrams to show the size effect on
the $\mathcal{MT}$ phase transition. One can see that
for $M=200$ that is larger than $M=50$ in the main text,
the $\mathcal{MT}$-symmetric region undergoes a considerable
shrinkage for both zero and finite magnetic field. Although the magnetic field
has a smaller effect on the critical phase boundary
compared with that for $M=50$,
it still results in a considerable reduction of the order parameter $d_{\text{HS}}$,
which indicates that
the magnetic field always increases
the number of states with real energies, consistent
with our discussion on the Onsager-Lifshitz quantization
in the long-wavelength limit.

The size effect implies
that no phase transition can occur as the system
is infinitely large in the $x$-direction.
However, realistic physical systems always
possess a finite size and the boundary condition
can also be continuously tuned
in certain artificial systems
such as the electrical circuits.
Therefore, the $\mathcal{MT}$ transition can be promisingly
achieved by experiments.

\section{VII. $\mathcal{MT}$-BREAKING BY NON-HERMITIAN COMPLEX HOPPING}
\begin{figure}[h]
\centering
\includegraphics[width=0.7\textwidth]{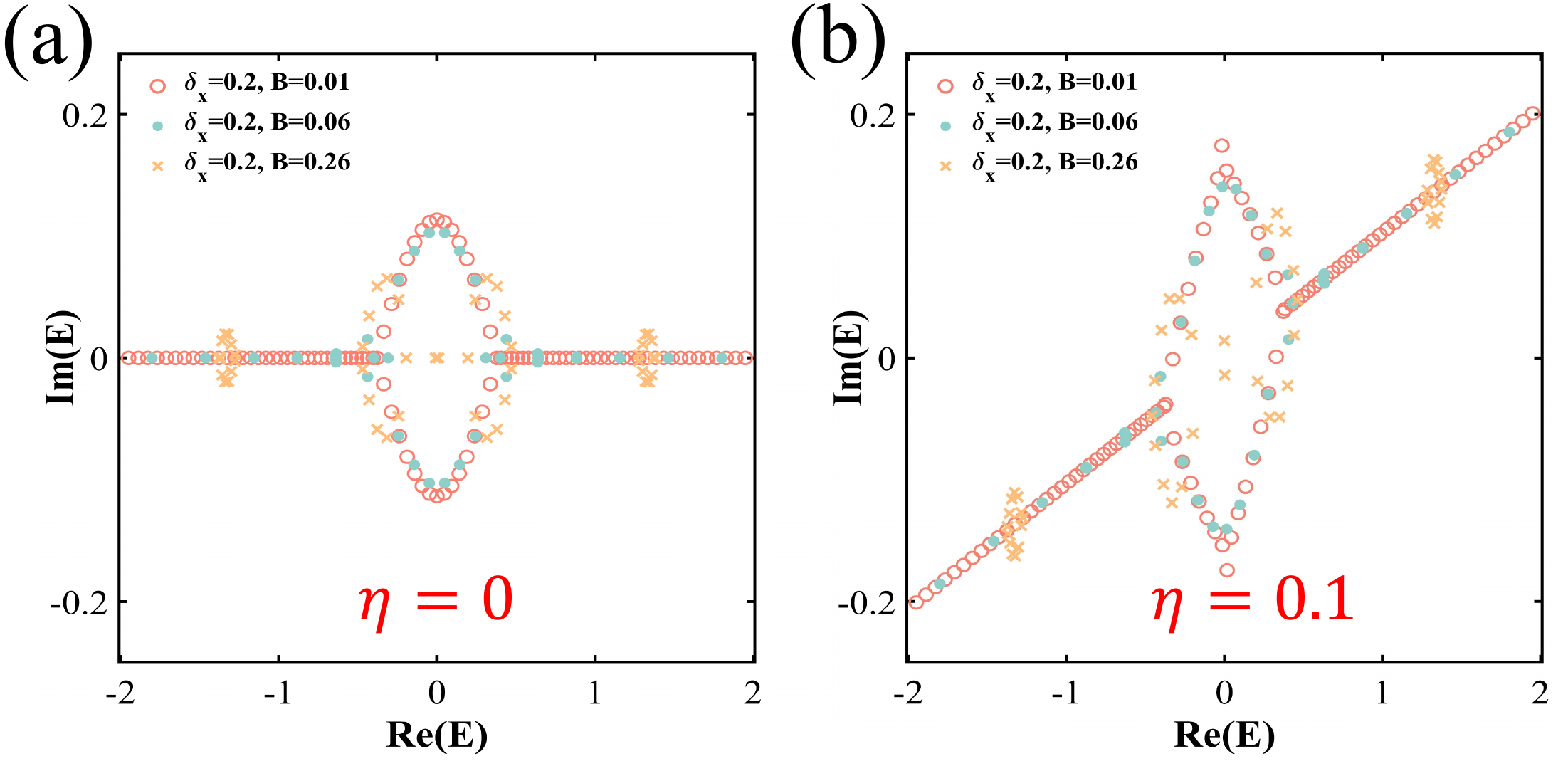}\\
\caption{Complex energy spectra under different $B$
with (a) $\eta=0$ and (b) $\eta=0.1$, respectively.
Common conditions: $x$-PBC, $y$-PBC, $t=0.5$ and $\delta_x=0.2$.}\label{SHopt1}
\end{figure}

In this section, we study the case that a small imaginary part $i\eta$ added to the hopping terms as $\tilde{t}_x^\pm=t+i\eta\pm\delta_x$,
which yields a non-Hermitian complex hopping. Mathematically, this is just a
substitution $t\rightarrow \tilde{t}=t+i\eta$ in the $x$-direction, which leads to entirely complex energy spectra
including the energy levels in the long-wavelength limit; see Fig.\ref{SHopt1} for comparison.
From the symmetry perspective, the additional $i\eta$ term breaks the $\mathcal{MT}$ symmetry.
As a result, the original $\mathcal{MT}$ spectral transition disappears, which further verifies
the $\mathcal{MT}$ scenario in our work.

From Fig.\ref{SHopt1}, one can see that the real and imaginary parts of the eigenvalues
exhibit a linear relation in the long-wavelength limit (band edges), which can be solved analytically
as
\begin{equation}\label{LL_Chop}
\begin{split}
\tilde{\varepsilon}_n=
(n+1/2)\hbar\tilde{\omega}+\delta_x^2/\tilde{t},
\end{split}
\end{equation}
where $\tilde{\omega}=qB/\sqrt{m_xm_y}$ becomes complex due to the complex effective mass $m_x=\hbar^2/(2\tilde{t}a^2)$ in the
$x$-direction. Although the eigenvalues $\tilde{\varepsilon}_n$ become complex, the quantization persists.
It is verified by the numerical results in Fig.~R.5, in which
the Landau fan structure of the real components still possesses equal level spacing and
a linear $B$-dependence.

\begin{figure}[h]
\centering
\includegraphics[width=0.7\textwidth]{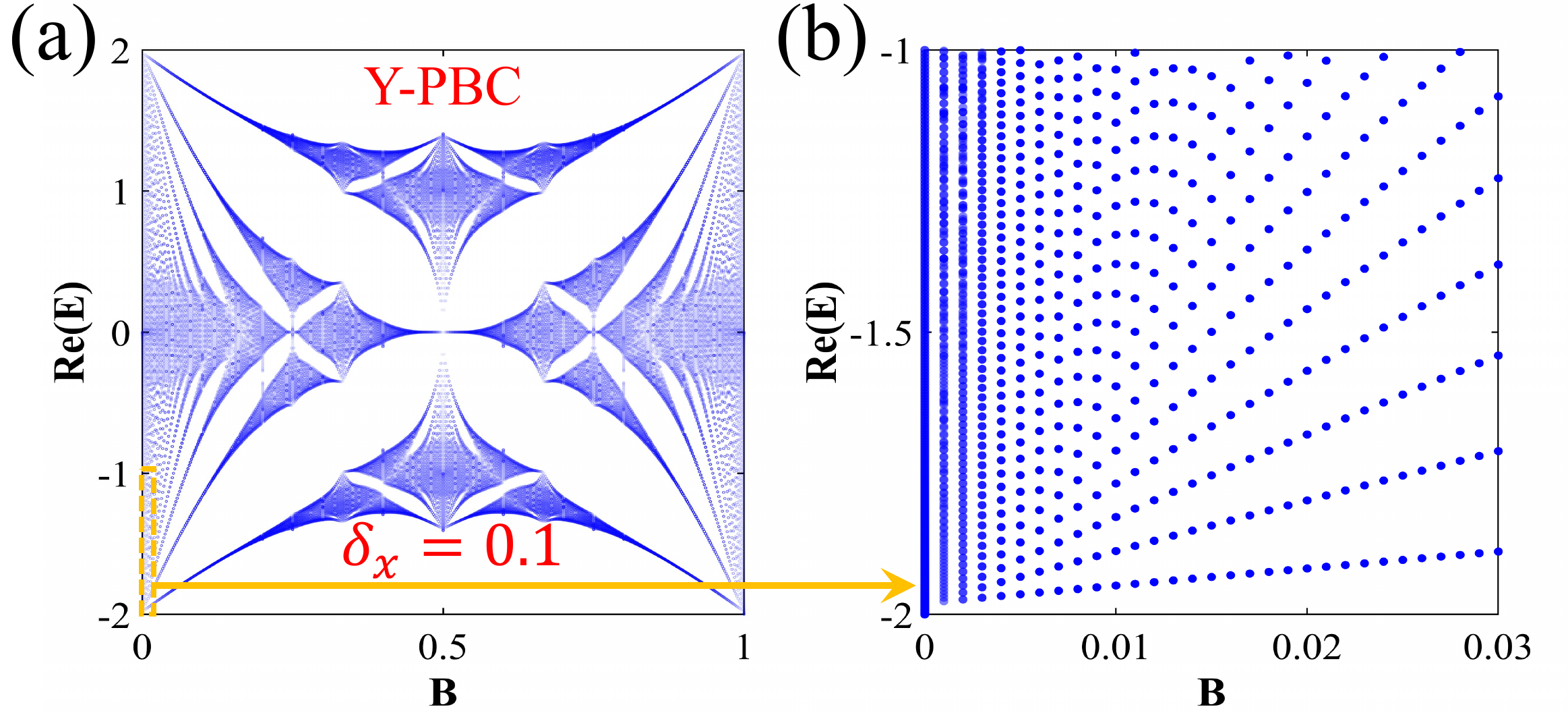}\\
\caption{Energy spectra under $x$-PBC, $y$-PBC with $\delta_x=0.1$, $\eta=0.1$, $t=0.5$, and $N=50$.}\label{SHopt2}
\end{figure}

\section{VIII. MAGNETIC ENERGY SPECTRA AND PHASE DIAGRAMS FOR NONRECIPROCAL
HONEYCOMB LATTICE}\label{NOTE10}
\begin{figure*}[!htbp]
\centering
\includegraphics[width=1\columnwidth]{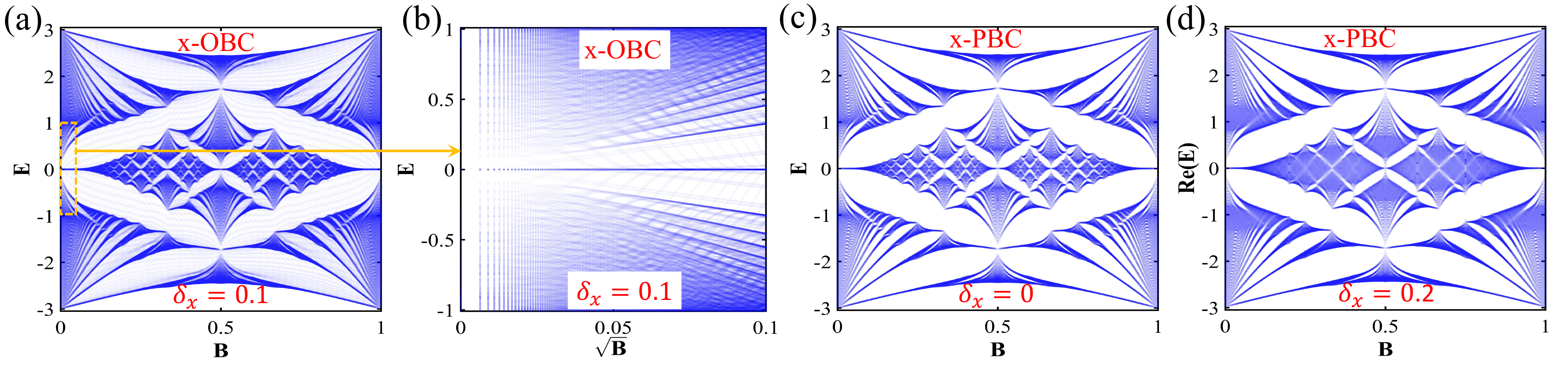}\\
\caption{(a) Energy spectra under the $x$-OBC and $y$-PBC with $\delta_x=0.1$ and $M=100$.
(b) Zoom of the Landau fan with the rescaled horizontal ordinate $\sqrt{B}$. Energy spectra under the
$x$-PBC and $y$-PBC calculated in the momentum space ($k_x, k_y$)
for (c) $\delta_x=0$ and (d) $\delta_x=0.2$. In all figures, $t=1$.}\label{fig7}
\end{figure*}

From Eq.~\eqref{Graphene_XYH} and the Fourier
transformed Hamiltonian $\tilde{H}'$, we plot the magnetic energy spectra in Fig.~\ref{fig7} under both
the $x$-OBC and $x$-PBC for the nonreciprocal honeycomb lattice. The spectra under the $x$-OBC are entirely
real despite the nonreciprocity. In contrast, complex spectra are
induced by the nonreciprocity under the $x$-PBC, where the fractal
patterns merge into continuous pieces
in the parametric regions far away from the long-wavelength limit.
In the vicinity of the Dirac points,
the same Landau fan structures arise for both the $x$-OBC and $x$-PBC;
see Figs.~\ref{fig7}(a) and \ref{fig7}(d). In particular,
the quantized energy levels satisfy
$E_n\propto\pm\sqrt{nB}$ with $n=0,1,\cdots$; see Fig.~\ref{fig7}(b),
manifesting the massless Dirac particle.

\begin{figure}[ht]
\centering
\includegraphics[width=0.7\columnwidth]{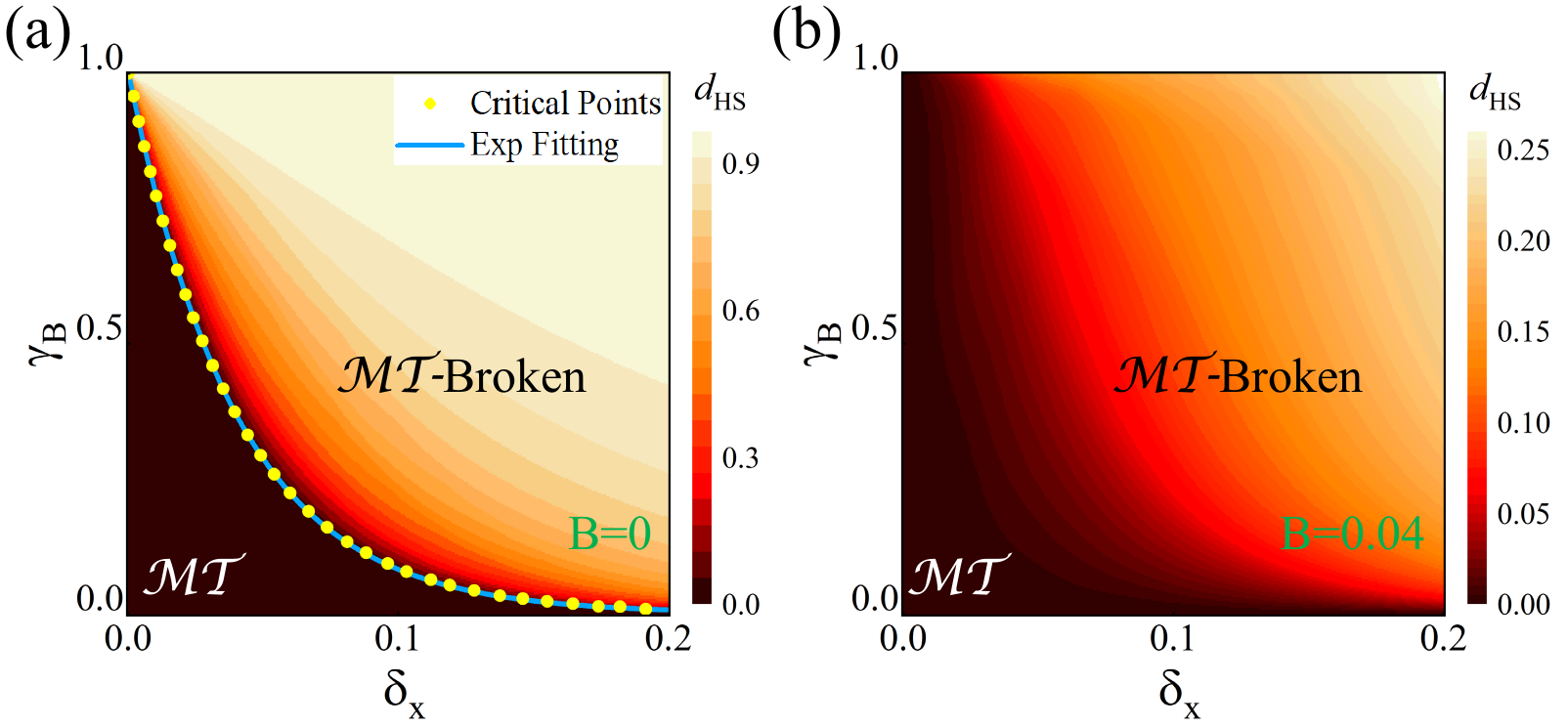}\\
\caption{The order parameter $d_{\text{HS}}$ as
a function of $\delta_x$ and $\gamma_B$ with (a) $B=0$ and (b) $B=0.04$.
Critical points in (a) mark the real-to-complex
spectral transition. The phase boundaries defined by
the $d_{\text{HS}}$ contours are fitted by the exponential functions.
Other parameters are set as $M=25$ and $t=1$.}\label{fig8}
\end{figure}

To study the $\mathcal{MT}$-symmetry breaking, we introduce a tunable boundary hopping, $\gamma_B(t_x^+ b^{\dagger}_{1,n}a_{M,n}+t_x^- a^{\dagger}_{M,n}b_{1,n})$,
between the outmost
sites $(1,n)_{\text{B}}$ and $(M,n)_{\text{A}}$,
where $\gamma_B\in[0,1]$ with
 $\gamma_B=0$
and $\gamma_B=1$ corresponding to the $x$-OBC and $x$-PBC,
respectively. We calculate the order parameter $d_{\text{HS}}$ using $\mathcal{MT}\psi^R_{a(b),i}(m,k_y)=\psi^{R*}_{a(b),i}(m,k_y)$
and plot the phase diagrams
in Fig.~\ref{fig8}.
The phase diagrams resemble the main text fig(4) for the square lattice quite well,
which reveals the universality of the spectral phase transition
induced by spontaneously $\mathcal{MT}$-symmetry breaking,
and a magnetic field
can effectively suppress
the $\mathcal{MT}$-symmetry breaking.

All the results in this section are consistent with those
of the nonreciprocal square lattice.


\begin{thebibliography}{70}%
\makeatletter
\providecommand \@ifxundefined [1]{%
 \@ifx{#1\undefined}
}%
\providecommand \@ifnum [1]{%
 \ifnum #1\expandafter \@firstoftwo
 \else \expandafter \@secondoftwo
 \fi
}%
\providecommand \@ifx [1]{%
 \ifx #1\expandafter \@firstoftwo
 \else \expandafter \@secondoftwo
 \fi
}%
\providecommand \natexlab [1]{#1}%
\providecommand \enquote  [1]{``#1''}%
\providecommand \bibnamefont  [1]{#1}%
\providecommand \bibfnamefont [1]{#1}%
\providecommand \citenamefont [1]{#1}%
\providecommand \href@noop [0]{\@secondoftwo}%
\providecommand \href [0]{\begingroup \@sanitize@url \@href}%
\providecommand \@href[1]{\@@startlink{#1}\@@href}%
\providecommand \@@href[1]{\endgroup#1\@@endlink}%
\providecommand \@sanitize@url [0]{\catcode `\\12\catcode `\$12\catcode
  `\&12\catcode `\#12\catcode `\^12\catcode `\_12\catcode `\%12\relax}%
\providecommand \@@startlink[1]{}%
\providecommand \@@endlink[0]{}%
\providecommand \url  [0]{\begingroup\@sanitize@url \@url }%
\providecommand \@url [1]{\endgroup\@href {#1}{\urlprefix }}%
\providecommand \urlprefix  [0]{URL }%
\providecommand \Eprint [0]{\href }%
\providecommand \doibase [0]{http://dx.doi.org/}%
\providecommand \selectlanguage [0]{\@gobble}%
\providecommand \bibinfo  [0]{\@secondoftwo}%
\providecommand \bibfield  [0]{\@secondoftwo}%
\providecommand \translation [1]{[#1]}%
\providecommand \BibitemOpen [0]{}%
\providecommand \bibitemStop [0]{}%
\providecommand \bibitemNoStop [0]{.\EOS\space}%
\providecommand \EOS [0]{\spacefactor3000\relax}%
\providecommand \BibitemShut  [1]{\csname bibitem#1\endcsname}%
\let\auto@bib@innerbib\@empty
\bibitem [{\citenamefont {Bender}\ and\ \citenamefont
  {Boettcher}(1998)}]{Bender98prl}%
  \BibitemOpen
  \bibfield  {author} {\bibinfo {author} {\bibfnamefont {C.~M.}\ \bibnamefont
  {Bender}}\ and\ \bibinfo {author} {\bibfnamefont {S.}~\bibnamefont
  {Boettcher}},\ }\href {\doibase 10.1103/PhysRevLett.80.5243} {\bibfield
  {journal} {\bibinfo  {journal} {Phys. Rev. Lett.}\ }\textbf {\bibinfo
  {volume} {80}},\ \bibinfo {pages} {5243} (\bibinfo {year}
  {1998})}\BibitemShut {NoStop}%
\bibitem [{\citenamefont {Bender}(2007)}]{bender2007making}%
  \BibitemOpen
  \bibfield  {author} {\bibinfo {author} {\bibfnamefont {C.~M.}\ \bibnamefont
  {Bender}},\ }\href
  {https://iopscience.iop.org/article/10.1088/0034-4885/70/6/R03} {\bibfield
  {journal} {\bibinfo  {journal} {Reports on Progress in Physics}\ }\textbf
  {\bibinfo {volume} {70}},\ \bibinfo {pages} {947} (\bibinfo {year}
  {2007})}\BibitemShut {NoStop}%
\bibitem [{\citenamefont {Ashida}\ \emph {et~al.}(2020)\citenamefont {Ashida},
  \citenamefont {Gong},\ and\ \citenamefont {Ueda}}]{ashida2020non}%
  \BibitemOpen
  \bibfield  {author} {\bibinfo {author} {\bibfnamefont {Y.}~\bibnamefont
  {Ashida}}, \bibinfo {author} {\bibfnamefont {Z.}~\bibnamefont {Gong}}, \ and\
  \bibinfo {author} {\bibfnamefont {M.}~\bibnamefont {Ueda}},\ }\href
  {https://www.tandfonline.com/doi/abs/10.1080/00018732.2021.1876991}
  {\bibfield  {journal} {\bibinfo  {journal} {Advances in Physics}\ }\textbf
  {\bibinfo {volume} {69}},\ \bibinfo {pages} {249} (\bibinfo {year}
  {2020})}\BibitemShut {NoStop}%
\bibitem [{\citenamefont {Feng}\ \emph {et~al.}(2017)\citenamefont {Feng},
  \citenamefont {El-Ganainy},\ and\ \citenamefont {Ge}}]{feng2017non}%
  \BibitemOpen
  \bibfield  {author} {\bibinfo {author} {\bibfnamefont {L.}~\bibnamefont
  {Feng}}, \bibinfo {author} {\bibfnamefont {R.}~\bibnamefont {El-Ganainy}}, \
  and\ \bibinfo {author} {\bibfnamefont {L.}~\bibnamefont {Ge}},\ }\href
  {https://www.nature.com/articles/s41566-017-0031-1} {\bibfield  {journal}
  {\bibinfo  {journal} {Nature Photonics}\ }\textbf {\bibinfo {volume} {11}},\
  \bibinfo {pages} {752} (\bibinfo {year} {2017})}\BibitemShut {NoStop}%
\bibitem [{\citenamefont {Ozawa}\ \emph {et~al.}(2019)\citenamefont {Ozawa},
  \citenamefont {Price}, \citenamefont {Amo}, \citenamefont {Goldman},
  \citenamefont {Hafezi}, \citenamefont {Lu}, \citenamefont {Rechtsman},
  \citenamefont {Schuster}, \citenamefont {Simon}, \citenamefont {Zilberberg},\
  and\ \citenamefont {Carusotto}}]{Ozawa2019rmp}%
  \BibitemOpen
  \bibfield  {author} {\bibinfo {author} {\bibfnamefont {T.}~\bibnamefont
  {Ozawa}}, \bibinfo {author} {\bibfnamefont {H.~M.}\ \bibnamefont {Price}},
  \bibinfo {author} {\bibfnamefont {A.}~\bibnamefont {Amo}}, \bibinfo {author}
  {\bibfnamefont {N.}~\bibnamefont {Goldman}}, \bibinfo {author} {\bibfnamefont
  {M.}~\bibnamefont {Hafezi}}, \bibinfo {author} {\bibfnamefont
  {L.}~\bibnamefont {Lu}}, \bibinfo {author} {\bibfnamefont {M.~C.}\
  \bibnamefont {Rechtsman}}, \bibinfo {author} {\bibfnamefont {D.}~\bibnamefont
  {Schuster}}, \bibinfo {author} {\bibfnamefont {J.}~\bibnamefont {Simon}},
  \bibinfo {author} {\bibfnamefont {O.}~\bibnamefont {Zilberberg}}, \ and\
  \bibinfo {author} {\bibfnamefont {I.}~\bibnamefont {Carusotto}},\ }\href
  {\doibase 10.1103/RevModPhys.91.015006} {\bibfield  {journal} {\bibinfo
  {journal} {Rev. Mod. Phys.}\ }\textbf {\bibinfo {volume} {91}},\ \bibinfo
  {pages} {015006} (\bibinfo {year} {2019})}\BibitemShut {NoStop}%
\bibitem [{\citenamefont {Chen}\ \emph {et~al.}(2017)\citenamefont {Chen},
  \citenamefont {{\"O}zdemir}, \citenamefont {Zhao}, \citenamefont {Wiersig},\
  and\ \citenamefont {Yang}}]{chen2017exceptional}%
  \BibitemOpen
  \bibfield  {author} {\bibinfo {author} {\bibfnamefont {W.}~\bibnamefont
  {Chen}}, \bibinfo {author} {\bibfnamefont {{\c{S}}.~K.}\ \bibnamefont
  {{\"O}zdemir}}, \bibinfo {author} {\bibfnamefont {G.}~\bibnamefont {Zhao}},
  \bibinfo {author} {\bibfnamefont {J.}~\bibnamefont {Wiersig}}, \ and\
  \bibinfo {author} {\bibfnamefont {L.}~\bibnamefont {Yang}},\ }\href@noop {}
  {\bibfield  {journal} {\bibinfo  {journal} {Nature}\ }\textbf {\bibinfo
  {volume} {548}},\ \bibinfo {pages} {192} (\bibinfo {year}
  {2017})}\BibitemShut {NoStop}%
\bibitem [{\citenamefont {El-Ganainy}\ \emph {et~al.}(2018)\citenamefont
  {El-Ganainy}, \citenamefont {Makris}, \citenamefont {Khajavikhan},
  \citenamefont {Musslimani}, \citenamefont {Rotter},\ and\ \citenamefont
  {Christodoulides}}]{el2018non}%
  \BibitemOpen
  \bibfield  {author} {\bibinfo {author} {\bibfnamefont {R.}~\bibnamefont
  {El-Ganainy}}, \bibinfo {author} {\bibfnamefont {K.~G.}\ \bibnamefont
  {Makris}}, \bibinfo {author} {\bibfnamefont {M.}~\bibnamefont {Khajavikhan}},
  \bibinfo {author} {\bibfnamefont {Z.~H.}\ \bibnamefont {Musslimani}},
  \bibinfo {author} {\bibfnamefont {S.}~\bibnamefont {Rotter}}, \ and\ \bibinfo
  {author} {\bibfnamefont {D.~N.}\ \bibnamefont {Christodoulides}},\ }\href
  {https://www.nature.com/articles/nphys4323} {\bibfield  {journal} {\bibinfo
  {journal} {Nature Physics}\ }\textbf {\bibinfo {volume} {14}},\ \bibinfo
  {pages} {11} (\bibinfo {year} {2018})}\BibitemShut {NoStop}%
\bibitem [{\citenamefont {{\"O}zdemir}\ \emph {et~al.}(2019)\citenamefont
  {{\"O}zdemir}, \citenamefont {Rotter}, \citenamefont {Nori},\ and\
  \citenamefont {Yang}}]{ozdemir2019parity}%
  \BibitemOpen
  \bibfield  {author} {\bibinfo {author} {\bibfnamefont {{\c{S}}.~K.}\
  \bibnamefont {{\"O}zdemir}}, \bibinfo {author} {\bibfnamefont
  {S.}~\bibnamefont {Rotter}}, \bibinfo {author} {\bibfnamefont
  {F.}~\bibnamefont {Nori}}, \ and\ \bibinfo {author} {\bibfnamefont
  {L.}~\bibnamefont {Yang}},\ }\href@noop {} {\bibfield  {journal} {\bibinfo
  {journal} {Nature materials}\ }\textbf {\bibinfo {volume} {18}},\ \bibinfo
  {pages} {783} (\bibinfo {year} {2019})}\BibitemShut {NoStop}%
\bibitem [{\citenamefont {Miri}\ and\ \citenamefont
  {Alu}(2019)}]{miri2019exceptional}%
  \BibitemOpen
  \bibfield  {author} {\bibinfo {author} {\bibfnamefont {M.-A.}\ \bibnamefont
  {Miri}}\ and\ \bibinfo {author} {\bibfnamefont {A.}~\bibnamefont {Alu}},\
  }\href
  {https://www.science.org/doi/full/10.1126/science.aar7709?casa_token=7YjnjFPGrvkAAAAA%3AY60Ty3BZ0Kv_X7rsQqNgrGF8KSU8IsPTPNUf9CGiF161UzYvpcSXQwlJ2EyzWsoUfhzEUtLt_NcC9g}
  {\bibfield  {journal} {\bibinfo  {journal} {Science}\ }\textbf {\bibinfo
  {volume} {363}},\ \bibinfo {pages} {42} (\bibinfo {year} {2019})}\BibitemShut
  {NoStop}%
\bibitem [{\citenamefont {Rotter}(1991)}]{rotter1991continuum}%
  \BibitemOpen
  \bibfield  {author} {\bibinfo {author} {\bibfnamefont {I.}~\bibnamefont
  {Rotter}},\ }\href
  {https://iopscience.iop.org/article/10.1088/0034-4885/54/4/003} {\bibfield
  {journal} {\bibinfo  {journal} {Reports on Progress in Physics}\ }\textbf
  {\bibinfo {volume} {54}},\ \bibinfo {pages} {635} (\bibinfo {year}
  {1991})}\BibitemShut {NoStop}%
\bibitem [{\citenamefont {Rotter}(2009)}]{rotter2009non}%
  \BibitemOpen
  \bibfield  {author} {\bibinfo {author} {\bibfnamefont {I.}~\bibnamefont
  {Rotter}},\ }\href
  {https://iopscience.iop.org/article/10.1088/1751-8113/42/15/153001}
  {\bibfield  {journal} {\bibinfo  {journal} {Journal of Physics A:
  Mathematical and Theoretical}\ }\textbf {\bibinfo {volume} {42}},\ \bibinfo
  {pages} {153001} (\bibinfo {year} {2009})}\BibitemShut {NoStop}%
\bibitem [{\citenamefont {Malzard}\ \emph {et~al.}(2015)\citenamefont
  {Malzard}, \citenamefont {Poli},\ and\ \citenamefont
  {Schomerus}}]{Malzard2015prl}%
  \BibitemOpen
  \bibfield  {author} {\bibinfo {author} {\bibfnamefont {S.}~\bibnamefont
  {Malzard}}, \bibinfo {author} {\bibfnamefont {C.}~\bibnamefont {Poli}}, \
  and\ \bibinfo {author} {\bibfnamefont {H.}~\bibnamefont {Schomerus}},\ }\href
  {\doibase 10.1103/PhysRevLett.115.200402} {\bibfield  {journal} {\bibinfo
  {journal} {Phys. Rev. Lett.}\ }\textbf {\bibinfo {volume} {115}},\ \bibinfo
  {pages} {200402} (\bibinfo {year} {2015})}\BibitemShut {NoStop}%
\bibitem [{\citenamefont {Diehl}\ \emph {et~al.}(2008)\citenamefont {Diehl},
  \citenamefont {Micheli}, \citenamefont {Kantian}, \citenamefont {Kraus},
  \citenamefont {B{\"u}chler},\ and\ \citenamefont
  {Zoller}}]{diehl2008quantum}%
  \BibitemOpen
  \bibfield  {author} {\bibinfo {author} {\bibfnamefont {S.}~\bibnamefont
  {Diehl}}, \bibinfo {author} {\bibfnamefont {A.}~\bibnamefont {Micheli}},
  \bibinfo {author} {\bibfnamefont {A.}~\bibnamefont {Kantian}}, \bibinfo
  {author} {\bibfnamefont {B.}~\bibnamefont {Kraus}}, \bibinfo {author}
  {\bibfnamefont {H.}~\bibnamefont {B{\"u}chler}}, \ and\ \bibinfo {author}
  {\bibfnamefont {P.}~\bibnamefont {Zoller}},\ }\href
  {https://www.nature.com/articles/nphys1073} {\bibfield  {journal} {\bibinfo
  {journal} {Nature Physics}\ }\textbf {\bibinfo {volume} {4}},\ \bibinfo
  {pages} {878} (\bibinfo {year} {2008})}\BibitemShut {NoStop}%
\bibitem [{\citenamefont {Kozii}\ and\ \citenamefont
  {Fu}(2017)}]{kozii2017nonhermitian}%
  \BibitemOpen
  \bibfield  {author} {\bibinfo {author} {\bibfnamefont {V.}~\bibnamefont
  {Kozii}}\ and\ \bibinfo {author} {\bibfnamefont {L.}~\bibnamefont {Fu}},\
  }\href {https://arxiv.org/abs/1708.05841} {\enquote {\bibinfo {title}
  {Non-hermitian topological theory of finite-lifetime quasiparticles:
  Prediction of bulk fermi arc due to exceptional point},}\ } (\bibinfo {year}
  {2017}),\ \Eprint {http://arxiv.org/abs/1708.05841} {arXiv:1708.05841
  [cond-mat.mes-hall]} \BibitemShut {NoStop}%
\bibitem [{\citenamefont {Shen}\ and\ \citenamefont {Fu}(2018)}]{shen2018prl}%
  \BibitemOpen
  \bibfield  {author} {\bibinfo {author} {\bibfnamefont {H.}~\bibnamefont
  {Shen}}\ and\ \bibinfo {author} {\bibfnamefont {L.}~\bibnamefont {Fu}},\
  }\href {\doibase 10.1103/PhysRevLett.121.026403} {\bibfield  {journal}
  {\bibinfo  {journal} {Phys. Rev. Lett.}\ }\textbf {\bibinfo {volume} {121}},\
  \bibinfo {pages} {026403} (\bibinfo {year} {2018})}\BibitemShut {NoStop}%
\bibitem [{\citenamefont {Papaj}\ \emph {et~al.}(2019)\citenamefont {Papaj},
  \citenamefont {Isobe},\ and\ \citenamefont {Fu}}]{Papaj2019prb}%
  \BibitemOpen
  \bibfield  {author} {\bibinfo {author} {\bibfnamefont {M.}~\bibnamefont
  {Papaj}}, \bibinfo {author} {\bibfnamefont {H.}~\bibnamefont {Isobe}}, \ and\
  \bibinfo {author} {\bibfnamefont {L.}~\bibnamefont {Fu}},\ }\href {\doibase
  10.1103/PhysRevB.99.201107} {\bibfield  {journal} {\bibinfo  {journal} {Phys.
  Rev. B}\ }\textbf {\bibinfo {volume} {99}},\ \bibinfo {pages} {201107}
  (\bibinfo {year} {2019})}\BibitemShut {NoStop}%
\bibitem [{\citenamefont {Yoshida}\ \emph {et~al.}(2018)\citenamefont
  {Yoshida}, \citenamefont {Peters},\ and\ \citenamefont
  {Kawakami}}]{Yoshida18prb}%
  \BibitemOpen
  \bibfield  {author} {\bibinfo {author} {\bibfnamefont {T.}~\bibnamefont
  {Yoshida}}, \bibinfo {author} {\bibfnamefont {R.}~\bibnamefont {Peters}}, \
  and\ \bibinfo {author} {\bibfnamefont {N.}~\bibnamefont {Kawakami}},\ }\href
  {\doibase 10.1103/PhysRevB.98.035141} {\bibfield  {journal} {\bibinfo
  {journal} {Phys. Rev. B}\ }\textbf {\bibinfo {volume} {98}},\ \bibinfo
  {pages} {035141} (\bibinfo {year} {2018})}\BibitemShut {NoStop}%
\bibitem [{\citenamefont {Schindler}\ \emph {et~al.}(2011)\citenamefont
  {Schindler}, \citenamefont {Li}, \citenamefont {Zheng}, \citenamefont
  {Ellis},\ and\ \citenamefont {Kottos}}]{Schindler2011pra}%
  \BibitemOpen
  \bibfield  {author} {\bibinfo {author} {\bibfnamefont {J.}~\bibnamefont
  {Schindler}}, \bibinfo {author} {\bibfnamefont {A.}~\bibnamefont {Li}},
  \bibinfo {author} {\bibfnamefont {M.~C.}\ \bibnamefont {Zheng}}, \bibinfo
  {author} {\bibfnamefont {F.~M.}\ \bibnamefont {Ellis}}, \ and\ \bibinfo
  {author} {\bibfnamefont {T.}~\bibnamefont {Kottos}},\ }\href {\doibase
  10.1103/PhysRevA.84.040101} {\bibfield  {journal} {\bibinfo  {journal} {Phys.
  Rev. A}\ }\textbf {\bibinfo {volume} {84}},\ \bibinfo {pages} {040101}
  (\bibinfo {year} {2011})}\BibitemShut {NoStop}%
\bibitem [{\citenamefont {Lee}\ \emph {et~al.}(2018)\citenamefont {Lee},
  \citenamefont {Imhof}, \citenamefont {Berger}, \citenamefont {Bayer},
  \citenamefont {Brehm}, \citenamefont {Molenkamp}, \citenamefont {Kiessling},\
  and\ \citenamefont {Thomale}}]{lee2018topolectrical}%
  \BibitemOpen
  \bibfield  {author} {\bibinfo {author} {\bibfnamefont {C.~H.}\ \bibnamefont
  {Lee}}, \bibinfo {author} {\bibfnamefont {S.}~\bibnamefont {Imhof}}, \bibinfo
  {author} {\bibfnamefont {C.}~\bibnamefont {Berger}}, \bibinfo {author}
  {\bibfnamefont {F.}~\bibnamefont {Bayer}}, \bibinfo {author} {\bibfnamefont
  {J.}~\bibnamefont {Brehm}}, \bibinfo {author} {\bibfnamefont {L.~W.}\
  \bibnamefont {Molenkamp}}, \bibinfo {author} {\bibfnamefont {T.}~\bibnamefont
  {Kiessling}}, \ and\ \bibinfo {author} {\bibfnamefont {R.}~\bibnamefont
  {Thomale}},\ }\href {https://www.nature.com/articles/s42005-018-0035-2}
  {\bibfield  {journal} {\bibinfo  {journal} {Communications Physics}\ }\textbf
  {\bibinfo {volume} {1}},\ \bibinfo {pages} {1} (\bibinfo {year}
  {2018})}\BibitemShut {NoStop}%
\bibitem [{\citenamefont {Kotwal}\ \emph {et~al.}(2021)\citenamefont {Kotwal},
  \citenamefont {Moseley}, \citenamefont {Stegmaier}, \citenamefont {Imhof},
  \citenamefont {Brand}, \citenamefont {Kie{\ss}ling}, \citenamefont {Thomale},
  \citenamefont {Ronellenfitsch},\ and\ \citenamefont
  {Dunkel}}]{kotwal2021active}%
  \BibitemOpen
  \bibfield  {author} {\bibinfo {author} {\bibfnamefont {T.}~\bibnamefont
  {Kotwal}}, \bibinfo {author} {\bibfnamefont {F.}~\bibnamefont {Moseley}},
  \bibinfo {author} {\bibfnamefont {A.}~\bibnamefont {Stegmaier}}, \bibinfo
  {author} {\bibfnamefont {S.}~\bibnamefont {Imhof}}, \bibinfo {author}
  {\bibfnamefont {H.}~\bibnamefont {Brand}}, \bibinfo {author} {\bibfnamefont
  {T.}~\bibnamefont {Kie{\ss}ling}}, \bibinfo {author} {\bibfnamefont
  {R.}~\bibnamefont {Thomale}}, \bibinfo {author} {\bibfnamefont
  {H.}~\bibnamefont {Ronellenfitsch}}, \ and\ \bibinfo {author} {\bibfnamefont
  {J.}~\bibnamefont {Dunkel}},\ }\href
  {https://www.pnas.org/doi/10.1073/pnas.2106411118} {\bibfield  {journal}
  {\bibinfo  {journal} {Proceedings of the National Academy of Sciences}\
  }\textbf {\bibinfo {volume} {118}} (\bibinfo {year} {2021})}\BibitemShut
  {NoStop}%
\bibitem [{\citenamefont {Ningyuan}\ \emph {et~al.}(2015)\citenamefont
  {Ningyuan}, \citenamefont {Owens}, \citenamefont {Sommer}, \citenamefont
  {Schuster},\ and\ \citenamefont {Simon}}]{Ningyuan15prx}%
  \BibitemOpen
  \bibfield  {author} {\bibinfo {author} {\bibfnamefont {J.}~\bibnamefont
  {Ningyuan}}, \bibinfo {author} {\bibfnamefont {C.}~\bibnamefont {Owens}},
  \bibinfo {author} {\bibfnamefont {A.}~\bibnamefont {Sommer}}, \bibinfo
  {author} {\bibfnamefont {D.}~\bibnamefont {Schuster}}, \ and\ \bibinfo
  {author} {\bibfnamefont {J.}~\bibnamefont {Simon}},\ }\href {\doibase
  10.1103/PhysRevX.5.021031} {\bibfield  {journal} {\bibinfo  {journal} {Phys.
  Rev. X}\ }\textbf {\bibinfo {volume} {5}},\ \bibinfo {pages} {021031}
  (\bibinfo {year} {2015})}\BibitemShut {NoStop}%
\bibitem [{\citenamefont {Imhof}\ \emph {et~al.}(2018)\citenamefont {Imhof},
  \citenamefont {Berger}, \citenamefont {Bayer}, \citenamefont {Brehm},
  \citenamefont {Molenkamp}, \citenamefont {Kiessling}, \citenamefont
  {Schindler}, \citenamefont {Lee}, \citenamefont {Greiter}, \citenamefont
  {Neupert} \emph {et~al.}}]{imhof2018topolectrical}%
  \BibitemOpen
  \bibfield  {author} {\bibinfo {author} {\bibfnamefont {S.}~\bibnamefont
  {Imhof}}, \bibinfo {author} {\bibfnamefont {C.}~\bibnamefont {Berger}},
  \bibinfo {author} {\bibfnamefont {F.}~\bibnamefont {Bayer}}, \bibinfo
  {author} {\bibfnamefont {J.}~\bibnamefont {Brehm}}, \bibinfo {author}
  {\bibfnamefont {L.~W.}\ \bibnamefont {Molenkamp}}, \bibinfo {author}
  {\bibfnamefont {T.}~\bibnamefont {Kiessling}}, \bibinfo {author}
  {\bibfnamefont {F.}~\bibnamefont {Schindler}}, \bibinfo {author}
  {\bibfnamefont {C.~H.}\ \bibnamefont {Lee}}, \bibinfo {author} {\bibfnamefont
  {M.}~\bibnamefont {Greiter}}, \bibinfo {author} {\bibfnamefont
  {T.}~\bibnamefont {Neupert}},  \emph {et~al.},\ }\href
  {https://www.nature.com/articles/s41567-018-0246-1} {\bibfield  {journal}
  {\bibinfo  {journal} {Nature Physics}\ }\textbf {\bibinfo {volume} {14}},\
  \bibinfo {pages} {925} (\bibinfo {year} {2018})}\BibitemShut {NoStop}%
\bibitem [{\citenamefont {Shen}\ \emph {et~al.}(2018)\citenamefont {Shen},
  \citenamefont {Zhen},\ and\ \citenamefont {Fu}}]{shen18prl2}%
  \BibitemOpen
  \bibfield  {author} {\bibinfo {author} {\bibfnamefont {H.}~\bibnamefont
  {Shen}}, \bibinfo {author} {\bibfnamefont {B.}~\bibnamefont {Zhen}}, \ and\
  \bibinfo {author} {\bibfnamefont {L.}~\bibnamefont {Fu}},\ }\href {\doibase
  10.1103/PhysRevLett.120.146402} {\bibfield  {journal} {\bibinfo  {journal}
  {Phys. Rev. Lett.}\ }\textbf {\bibinfo {volume} {120}},\ \bibinfo {pages}
  {146402} (\bibinfo {year} {2018})}\BibitemShut {NoStop}%
\bibitem [{\citenamefont {Bergholtz}\ \emph {et~al.}(2021)\citenamefont
  {Bergholtz}, \citenamefont {Budich},\ and\ \citenamefont
  {Kunst}}]{Bergholtz21rmp}%
  \BibitemOpen
  \bibfield  {author} {\bibinfo {author} {\bibfnamefont {E.~J.}\ \bibnamefont
  {Bergholtz}}, \bibinfo {author} {\bibfnamefont {J.~C.}\ \bibnamefont
  {Budich}}, \ and\ \bibinfo {author} {\bibfnamefont {F.~K.}\ \bibnamefont
  {Kunst}},\ }\href {\doibase 10.1103/RevModPhys.93.015005} {\bibfield
  {journal} {\bibinfo  {journal} {Rev. Mod. Phys.}\ }\textbf {\bibinfo {volume}
  {93}},\ \bibinfo {pages} {015005} (\bibinfo {year} {2021})}\BibitemShut
  {NoStop}%
\bibitem [{\citenamefont {Lee}(2016)}]{Lee2016prl}%
  \BibitemOpen
  \bibfield  {author} {\bibinfo {author} {\bibfnamefont {T.~E.}\ \bibnamefont
  {Lee}},\ }\href {\doibase 10.1103/PhysRevLett.116.133903} {\bibfield
  {journal} {\bibinfo  {journal} {Phys. Rev. Lett.}\ }\textbf {\bibinfo
  {volume} {116}},\ \bibinfo {pages} {133903} (\bibinfo {year}
  {2016})}\BibitemShut {NoStop}%
\bibitem [{\citenamefont {Xiong}(2018)}]{xiong2018does}%
  \BibitemOpen
  \bibfield  {author} {\bibinfo {author} {\bibfnamefont {Y.}~\bibnamefont
  {Xiong}},\ }\href
  {https://iopscience.iop.org/article/10.1088/2399-6528/aab64a/meta} {\bibfield
   {journal} {\bibinfo  {journal} {Journal of Physics Communications}\ }\textbf
  {\bibinfo {volume} {2}},\ \bibinfo {pages} {035043} (\bibinfo {year}
  {2018})}\BibitemShut {NoStop}%
\bibitem [{\citenamefont {Gong}\ \emph {et~al.}(2018)\citenamefont {Gong},
  \citenamefont {Ashida}, \citenamefont {Kawabata}, \citenamefont {Takasan},
  \citenamefont {Higashikawa},\ and\ \citenamefont {Ueda}}]{Gong2018prx}%
  \BibitemOpen
  \bibfield  {author} {\bibinfo {author} {\bibfnamefont {Z.}~\bibnamefont
  {Gong}}, \bibinfo {author} {\bibfnamefont {Y.}~\bibnamefont {Ashida}},
  \bibinfo {author} {\bibfnamefont {K.}~\bibnamefont {Kawabata}}, \bibinfo
  {author} {\bibfnamefont {K.}~\bibnamefont {Takasan}}, \bibinfo {author}
  {\bibfnamefont {S.}~\bibnamefont {Higashikawa}}, \ and\ \bibinfo {author}
  {\bibfnamefont {M.}~\bibnamefont {Ueda}},\ }\href {\doibase
  10.1103/PhysRevX.8.031079} {\bibfield  {journal} {\bibinfo  {journal} {Phys.
  Rev. X}\ }\textbf {\bibinfo {volume} {8}},\ \bibinfo {pages} {031079}
  (\bibinfo {year} {2018})}\BibitemShut {NoStop}%
\bibitem [{\citenamefont {Kawabata}\ \emph {et~al.}(2019)\citenamefont
  {Kawabata}, \citenamefont {Higashikawa}, \citenamefont {Gong}, \citenamefont
  {Ashida},\ and\ \citenamefont {Ueda}}]{kawabata19nc}%
  \BibitemOpen
  \bibfield  {author} {\bibinfo {author} {\bibfnamefont {K.}~\bibnamefont
  {Kawabata}}, \bibinfo {author} {\bibfnamefont {S.}~\bibnamefont
  {Higashikawa}}, \bibinfo {author} {\bibfnamefont {Z.}~\bibnamefont {Gong}},
  \bibinfo {author} {\bibfnamefont {Y.}~\bibnamefont {Ashida}}, \ and\ \bibinfo
  {author} {\bibfnamefont {M.}~\bibnamefont {Ueda}},\ }\href
  {https://www.nature.com/articles/s41467-018-08254-y} {\bibfield  {journal}
  {\bibinfo  {journal} {Nature communications}\ }\textbf {\bibinfo {volume}
  {10}},\ \bibinfo {pages} {1} (\bibinfo {year} {2019})}\BibitemShut {NoStop}%
\bibitem [{\citenamefont {Xu}\ \emph {et~al.}(2017)\citenamefont {Xu},
  \citenamefont {Wang},\ and\ \citenamefont {Duan}}]{Xu2017prl}%
  \BibitemOpen
  \bibfield  {author} {\bibinfo {author} {\bibfnamefont {Y.}~\bibnamefont
  {Xu}}, \bibinfo {author} {\bibfnamefont {S.-T.}\ \bibnamefont {Wang}}, \ and\
  \bibinfo {author} {\bibfnamefont {L.-M.}\ \bibnamefont {Duan}},\ }\href
  {\doibase 10.1103/PhysRevLett.118.045701} {\bibfield  {journal} {\bibinfo
  {journal} {Phys. Rev. Lett.}\ }\textbf {\bibinfo {volume} {118}},\ \bibinfo
  {pages} {045701} (\bibinfo {year} {2017})}\BibitemShut {NoStop}%
\bibitem [{\citenamefont {Carlstr\"om}\ \emph {et~al.}(2019)\citenamefont
  {Carlstr\"om}, \citenamefont {St\aa{}lhammar}, \citenamefont {Budich},\ and\
  \citenamefont {Bergholtz}}]{Carlstr19prb}%
  \BibitemOpen
  \bibfield  {author} {\bibinfo {author} {\bibfnamefont {J.}~\bibnamefont
  {Carlstr\"om}}, \bibinfo {author} {\bibfnamefont {M.}~\bibnamefont
  {St\aa{}lhammar}}, \bibinfo {author} {\bibfnamefont {J.~C.}\ \bibnamefont
  {Budich}}, \ and\ \bibinfo {author} {\bibfnamefont {E.~J.}\ \bibnamefont
  {Bergholtz}},\ }\href {\doibase 10.1103/PhysRevB.99.161115} {\bibfield
  {journal} {\bibinfo  {journal} {Phys. Rev. B}\ }\textbf {\bibinfo {volume}
  {99}},\ \bibinfo {pages} {161115} (\bibinfo {year} {2019})}\BibitemShut
  {NoStop}%
\bibitem [{\citenamefont {Peng}\ \emph {et~al.}(2014)\citenamefont {Peng},
  \citenamefont {{\"O}zdemir}, \citenamefont {Lei}, \citenamefont {Monifi},
  \citenamefont {Gianfreda}, \citenamefont {Long}, \citenamefont {Fan},
  \citenamefont {Nori}, \citenamefont {Bender},\ and\ \citenamefont
  {Yang}}]{peng2014parity}%
  \BibitemOpen
  \bibfield  {author} {\bibinfo {author} {\bibfnamefont {B.}~\bibnamefont
  {Peng}}, \bibinfo {author} {\bibfnamefont {{\c{S}}.~K.}\ \bibnamefont
  {{\"O}zdemir}}, \bibinfo {author} {\bibfnamefont {F.}~\bibnamefont {Lei}},
  \bibinfo {author} {\bibfnamefont {F.}~\bibnamefont {Monifi}}, \bibinfo
  {author} {\bibfnamefont {M.}~\bibnamefont {Gianfreda}}, \bibinfo {author}
  {\bibfnamefont {G.~L.}\ \bibnamefont {Long}}, \bibinfo {author}
  {\bibfnamefont {S.}~\bibnamefont {Fan}}, \bibinfo {author} {\bibfnamefont
  {F.}~\bibnamefont {Nori}}, \bibinfo {author} {\bibfnamefont {C.~M.}\
  \bibnamefont {Bender}}, \ and\ \bibinfo {author} {\bibfnamefont
  {L.}~\bibnamefont {Yang}},\ }\href@noop {} {\bibfield  {journal} {\bibinfo
  {journal} {Nature Physics}\ }\textbf {\bibinfo {volume} {10}},\ \bibinfo
  {pages} {394} (\bibinfo {year} {2014})}\BibitemShut {NoStop}%
\bibitem [{\citenamefont {St-Jean}\ \emph {et~al.}(2017)\citenamefont
  {St-Jean}, \citenamefont {Goblot}, \citenamefont {Galopin}, \citenamefont
  {Lema{\^\i}tre}, \citenamefont {Ozawa}, \citenamefont {Le~Gratiet},
  \citenamefont {Sagnes}, \citenamefont {Bloch},\ and\ \citenamefont
  {Amo}}]{st2017lasing}%
  \BibitemOpen
  \bibfield  {author} {\bibinfo {author} {\bibfnamefont {P.}~\bibnamefont
  {St-Jean}}, \bibinfo {author} {\bibfnamefont {V.}~\bibnamefont {Goblot}},
  \bibinfo {author} {\bibfnamefont {E.}~\bibnamefont {Galopin}}, \bibinfo
  {author} {\bibfnamefont {A.}~\bibnamefont {Lema{\^\i}tre}}, \bibinfo {author}
  {\bibfnamefont {T.}~\bibnamefont {Ozawa}}, \bibinfo {author} {\bibfnamefont
  {L.}~\bibnamefont {Le~Gratiet}}, \bibinfo {author} {\bibfnamefont
  {I.}~\bibnamefont {Sagnes}}, \bibinfo {author} {\bibfnamefont
  {J.}~\bibnamefont {Bloch}}, \ and\ \bibinfo {author} {\bibfnamefont
  {A.}~\bibnamefont {Amo}},\ }\href
  {https://www.nature.com/articles/s41566-017-0006-2} {\bibfield  {journal}
  {\bibinfo  {journal} {Nature Photonics}\ }\textbf {\bibinfo {volume} {11}},\
  \bibinfo {pages} {651} (\bibinfo {year} {2017})}\BibitemShut {NoStop}%
\bibitem [{\citenamefont {Parto}\ \emph {et~al.}(2018)\citenamefont {Parto},
  \citenamefont {Wittek}, \citenamefont {Hodaei}, \citenamefont {Harari},
  \citenamefont {Bandres}, \citenamefont {Ren}, \citenamefont {Rechtsman},
  \citenamefont {Segev}, \citenamefont {Christodoulides},\ and\ \citenamefont
  {Khajavikhan}}]{Parto18prl}%
  \BibitemOpen
  \bibfield  {author} {\bibinfo {author} {\bibfnamefont {M.}~\bibnamefont
  {Parto}}, \bibinfo {author} {\bibfnamefont {S.}~\bibnamefont {Wittek}},
  \bibinfo {author} {\bibfnamefont {H.}~\bibnamefont {Hodaei}}, \bibinfo
  {author} {\bibfnamefont {G.}~\bibnamefont {Harari}}, \bibinfo {author}
  {\bibfnamefont {M.~A.}\ \bibnamefont {Bandres}}, \bibinfo {author}
  {\bibfnamefont {J.}~\bibnamefont {Ren}}, \bibinfo {author} {\bibfnamefont
  {M.~C.}\ \bibnamefont {Rechtsman}}, \bibinfo {author} {\bibfnamefont
  {M.}~\bibnamefont {Segev}}, \bibinfo {author} {\bibfnamefont {D.~N.}\
  \bibnamefont {Christodoulides}}, \ and\ \bibinfo {author} {\bibfnamefont
  {M.}~\bibnamefont {Khajavikhan}},\ }\href {\doibase
  10.1103/PhysRevLett.120.113901} {\bibfield  {journal} {\bibinfo  {journal}
  {Phys. Rev. Lett.}\ }\textbf {\bibinfo {volume} {120}},\ \bibinfo {pages}
  {113901} (\bibinfo {year} {2018})}\BibitemShut {NoStop}%
\bibitem [{\citenamefont {Yao}\ and\ \citenamefont
  {Wang}(2018)}]{Yaoshunyu2018prl}%
  \BibitemOpen
  \bibfield  {author} {\bibinfo {author} {\bibfnamefont {S.}~\bibnamefont
  {Yao}}\ and\ \bibinfo {author} {\bibfnamefont {Z.}~\bibnamefont {Wang}},\
  }\href {\doibase 10.1103/PhysRevLett.121.086803} {\bibfield  {journal}
  {\bibinfo  {journal} {Phys. Rev. Lett.}\ }\textbf {\bibinfo {volume} {121}},\
  \bibinfo {pages} {086803} (\bibinfo {year} {2018})}\BibitemShut {NoStop}%
\bibitem [{\citenamefont {Kunst}\ \emph {et~al.}(2018)\citenamefont {Kunst},
  \citenamefont {Edvardsson}, \citenamefont {Budich},\ and\ \citenamefont
  {Bergholtz}}]{Kunst2018prl}%
  \BibitemOpen
  \bibfield  {author} {\bibinfo {author} {\bibfnamefont {F.~K.}\ \bibnamefont
  {Kunst}}, \bibinfo {author} {\bibfnamefont {E.}~\bibnamefont {Edvardsson}},
  \bibinfo {author} {\bibfnamefont {J.~C.}\ \bibnamefont {Budich}}, \ and\
  \bibinfo {author} {\bibfnamefont {E.~J.}\ \bibnamefont {Bergholtz}},\ }\href
  {\doibase 10.1103/PhysRevLett.121.026808} {\bibfield  {journal} {\bibinfo
  {journal} {Phys. Rev. Lett.}\ }\textbf {\bibinfo {volume} {121}},\ \bibinfo
  {pages} {026808} (\bibinfo {year} {2018})}\BibitemShut {NoStop}%
\bibitem [{\citenamefont {Yokomizo}\ and\ \citenamefont
  {Murakami}(2019)}]{Yokomizo2019prl}%
  \BibitemOpen
  \bibfield  {author} {\bibinfo {author} {\bibfnamefont {K.}~\bibnamefont
  {Yokomizo}}\ and\ \bibinfo {author} {\bibfnamefont {S.}~\bibnamefont
  {Murakami}},\ }\href {\doibase 10.1103/PhysRevLett.123.066404} {\bibfield
  {journal} {\bibinfo  {journal} {Phys. Rev. Lett.}\ }\textbf {\bibinfo
  {volume} {123}},\ \bibinfo {pages} {066404} (\bibinfo {year}
  {2019})}\BibitemShut {NoStop}%
\bibitem [{\citenamefont {Borgnia}\ \emph {et~al.}(2020)\citenamefont
  {Borgnia}, \citenamefont {Kruchkov},\ and\ \citenamefont
  {Slager}}]{Borgnia20prl}%
  \BibitemOpen
  \bibfield  {author} {\bibinfo {author} {\bibfnamefont {D.~S.}\ \bibnamefont
  {Borgnia}}, \bibinfo {author} {\bibfnamefont {A.~J.}\ \bibnamefont
  {Kruchkov}}, \ and\ \bibinfo {author} {\bibfnamefont {R.-J.}\ \bibnamefont
  {Slager}},\ }\href {\doibase 10.1103/PhysRevLett.124.056802} {\bibfield
  {journal} {\bibinfo  {journal} {Phys. Rev. Lett.}\ }\textbf {\bibinfo
  {volume} {124}},\ \bibinfo {pages} {056802} (\bibinfo {year}
  {2020})}\BibitemShut {NoStop}%
\bibitem [{\citenamefont {Zhang}\ \emph {et~al.}(2020)\citenamefont {Zhang},
  \citenamefont {Yang},\ and\ \citenamefont {Fang}}]{Zhangkai2020prl}%
  \BibitemOpen
  \bibfield  {author} {\bibinfo {author} {\bibfnamefont {K.}~\bibnamefont
  {Zhang}}, \bibinfo {author} {\bibfnamefont {Z.}~\bibnamefont {Yang}}, \ and\
  \bibinfo {author} {\bibfnamefont {C.}~\bibnamefont {Fang}},\ }\href {\doibase
  10.1103/PhysRevLett.125.126402} {\bibfield  {journal} {\bibinfo  {journal}
  {Phys. Rev. Lett.}\ }\textbf {\bibinfo {volume} {125}},\ \bibinfo {pages}
  {126402} (\bibinfo {year} {2020})}\BibitemShut {NoStop}%
\bibitem [{\citenamefont {Okuma}\ \emph {et~al.}(2020)\citenamefont {Okuma},
  \citenamefont {Kawabata}, \citenamefont {Shiozaki},\ and\ \citenamefont
  {Sato}}]{Okuma2020prl}%
  \BibitemOpen
  \bibfield  {author} {\bibinfo {author} {\bibfnamefont {N.}~\bibnamefont
  {Okuma}}, \bibinfo {author} {\bibfnamefont {K.}~\bibnamefont {Kawabata}},
  \bibinfo {author} {\bibfnamefont {K.}~\bibnamefont {Shiozaki}}, \ and\
  \bibinfo {author} {\bibfnamefont {M.}~\bibnamefont {Sato}},\ }\href {\doibase
  10.1103/PhysRevLett.124.086801} {\bibfield  {journal} {\bibinfo  {journal}
  {Phys. Rev. Lett.}\ }\textbf {\bibinfo {volume} {124}},\ \bibinfo {pages}
  {086801} (\bibinfo {year} {2020})}\BibitemShut {NoStop}%
\bibitem [{\citenamefont {Yang}\ \emph {et~al.}(2020)\citenamefont {Yang},
  \citenamefont {Zhang}, \citenamefont {Fang},\ and\ \citenamefont
  {Hu}}]{Yangzhesen2020prl}%
  \BibitemOpen
  \bibfield  {author} {\bibinfo {author} {\bibfnamefont {Z.}~\bibnamefont
  {Yang}}, \bibinfo {author} {\bibfnamefont {K.}~\bibnamefont {Zhang}},
  \bibinfo {author} {\bibfnamefont {C.}~\bibnamefont {Fang}}, \ and\ \bibinfo
  {author} {\bibfnamefont {J.}~\bibnamefont {Hu}},\ }\href {\doibase
  10.1103/PhysRevLett.125.226402} {\bibfield  {journal} {\bibinfo  {journal}
  {Phys. Rev. Lett.}\ }\textbf {\bibinfo {volume} {125}},\ \bibinfo {pages}
  {226402} (\bibinfo {year} {2020})}\BibitemShut {NoStop}%
\bibitem [{\citenamefont {Li}\ \emph {et~al.}(2020)\citenamefont {Li},
  \citenamefont {Lee}, \citenamefont {Mu},\ and\ \citenamefont
  {Gong}}]{li2020critical}%
  \BibitemOpen
  \bibfield  {author} {\bibinfo {author} {\bibfnamefont {L.}~\bibnamefont
  {Li}}, \bibinfo {author} {\bibfnamefont {C.~H.}\ \bibnamefont {Lee}},
  \bibinfo {author} {\bibfnamefont {S.}~\bibnamefont {Mu}}, \ and\ \bibinfo
  {author} {\bibfnamefont {J.}~\bibnamefont {Gong}},\ }\href
  {https://www.nature.com/articles/s41467-020-18917-4} {\bibfield  {journal}
  {\bibinfo  {journal} {Nature communications}\ }\textbf {\bibinfo {volume}
  {11}},\ \bibinfo {pages} {1} (\bibinfo {year} {2020})}\BibitemShut {NoStop}%
\bibitem [{\citenamefont {Helbig}\ \emph {et~al.}(2020)\citenamefont {Helbig},
  \citenamefont {Hofmann}, \citenamefont {Imhof}, \citenamefont {Abdelghany},
  \citenamefont {Kiessling}, \citenamefont {Molenkamp}, \citenamefont {Lee},
  \citenamefont {Szameit}, \citenamefont {Greiter},\ and\ \citenamefont
  {Thomale}}]{helbig2020generalized}%
  \BibitemOpen
  \bibfield  {author} {\bibinfo {author} {\bibfnamefont {T.}~\bibnamefont
  {Helbig}}, \bibinfo {author} {\bibfnamefont {T.}~\bibnamefont {Hofmann}},
  \bibinfo {author} {\bibfnamefont {S.}~\bibnamefont {Imhof}}, \bibinfo
  {author} {\bibfnamefont {M.}~\bibnamefont {Abdelghany}}, \bibinfo {author}
  {\bibfnamefont {T.}~\bibnamefont {Kiessling}}, \bibinfo {author}
  {\bibfnamefont {L.}~\bibnamefont {Molenkamp}}, \bibinfo {author}
  {\bibfnamefont {C.}~\bibnamefont {Lee}}, \bibinfo {author} {\bibfnamefont
  {A.}~\bibnamefont {Szameit}}, \bibinfo {author} {\bibfnamefont
  {M.}~\bibnamefont {Greiter}}, \ and\ \bibinfo {author} {\bibfnamefont
  {R.}~\bibnamefont {Thomale}},\ }\href
  {https://www.nature.com/articles/s41567-020-0922-9} {\bibfield  {journal}
  {\bibinfo  {journal} {Nature Physics}\ }\textbf {\bibinfo {volume} {16}},\
  \bibinfo {pages} {747} (\bibinfo {year} {2020})}\BibitemShut {NoStop}%
\bibitem [{\citenamefont {Xiao}\ \emph {et~al.}(2020)\citenamefont {Xiao},
  \citenamefont {Deng}, \citenamefont {Wang}, \citenamefont {Zhu},
  \citenamefont {Wang}, \citenamefont {Yi},\ and\ \citenamefont
  {Xue}}]{xiao2020non}%
  \BibitemOpen
  \bibfield  {author} {\bibinfo {author} {\bibfnamefont {L.}~\bibnamefont
  {Xiao}}, \bibinfo {author} {\bibfnamefont {T.}~\bibnamefont {Deng}}, \bibinfo
  {author} {\bibfnamefont {K.}~\bibnamefont {Wang}}, \bibinfo {author}
  {\bibfnamefont {G.}~\bibnamefont {Zhu}}, \bibinfo {author} {\bibfnamefont
  {Z.}~\bibnamefont {Wang}}, \bibinfo {author} {\bibfnamefont {W.}~\bibnamefont
  {Yi}}, \ and\ \bibinfo {author} {\bibfnamefont {P.}~\bibnamefont {Xue}},\
  }\href {https://www.nature.com/articles/s41567-020-0836-6} {\bibfield
  {journal} {\bibinfo  {journal} {Nature Physics}\ }\textbf {\bibinfo {volume}
  {16}},\ \bibinfo {pages} {761} (\bibinfo {year} {2020})}\BibitemShut
  {NoStop}%
\bibitem [{\citenamefont {Weidemann}\ \emph {et~al.}(2020)\citenamefont
  {Weidemann}, \citenamefont {Kremer}, \citenamefont {Helbig}, \citenamefont
  {Hofmann}, \citenamefont {Stegmaier}, \citenamefont {Greiter}, \citenamefont
  {Thomale},\ and\ \citenamefont {Szameit}}]{weidemann2020topological}%
  \BibitemOpen
  \bibfield  {author} {\bibinfo {author} {\bibfnamefont {S.}~\bibnamefont
  {Weidemann}}, \bibinfo {author} {\bibfnamefont {M.}~\bibnamefont {Kremer}},
  \bibinfo {author} {\bibfnamefont {T.}~\bibnamefont {Helbig}}, \bibinfo
  {author} {\bibfnamefont {T.}~\bibnamefont {Hofmann}}, \bibinfo {author}
  {\bibfnamefont {A.}~\bibnamefont {Stegmaier}}, \bibinfo {author}
  {\bibfnamefont {M.}~\bibnamefont {Greiter}}, \bibinfo {author} {\bibfnamefont
  {R.}~\bibnamefont {Thomale}}, \ and\ \bibinfo {author} {\bibfnamefont
  {A.}~\bibnamefont {Szameit}},\ }\href
  {https://www.science.org/doi/full/10.1126/science.aaz8727?casa_token=a6tJwzkgkvcAAAAA%3Aa2XPkVUQUk_w2hvmnifH4QJRNFrZ5_ZRdUMh7JNS34SE2j-2z896EFxd7sGB0CC1l-DusKtw6ugSIg}
  {\bibfield  {journal} {\bibinfo  {journal} {Science}\ }\textbf {\bibinfo
  {volume} {368}},\ \bibinfo {pages} {311} (\bibinfo {year}
  {2020})}\BibitemShut {NoStop}%
\bibitem [{\citenamefont {Ghatak}\ \emph {et~al.}(2020)\citenamefont {Ghatak},
  \citenamefont {Brandenbourger}, \citenamefont {van Wezel},\ and\
  \citenamefont {Coulais}}]{ghatak2020observation}%
  \BibitemOpen
  \bibfield  {author} {\bibinfo {author} {\bibfnamefont {A.}~\bibnamefont
  {Ghatak}}, \bibinfo {author} {\bibfnamefont {M.}~\bibnamefont
  {Brandenbourger}}, \bibinfo {author} {\bibfnamefont {J.}~\bibnamefont {van
  Wezel}}, \ and\ \bibinfo {author} {\bibfnamefont {C.}~\bibnamefont
  {Coulais}},\ }\href {https://www.pnas.org/doi/abs/10.1073/pnas.2010580117}
  {\bibfield  {journal} {\bibinfo  {journal} {Proceedings of the National
  Academy of Sciences}\ }\textbf {\bibinfo {volume} {117}},\ \bibinfo {pages}
  {29561} (\bibinfo {year} {2020})}\BibitemShut {NoStop}%
\bibitem [{\citenamefont {Palacios}\ \emph {et~al.}(2021)\citenamefont
  {Palacios}, \citenamefont {Tchoumakov}, \citenamefont {Guix}, \citenamefont
  {Pagonabarraga}, \citenamefont {S{\'a}nchez},\ and\ \citenamefont
  {G~Grushin}}]{palacios2021guided}%
  \BibitemOpen
  \bibfield  {author} {\bibinfo {author} {\bibfnamefont {L.~S.}\ \bibnamefont
  {Palacios}}, \bibinfo {author} {\bibfnamefont {S.}~\bibnamefont
  {Tchoumakov}}, \bibinfo {author} {\bibfnamefont {M.}~\bibnamefont {Guix}},
  \bibinfo {author} {\bibfnamefont {I.}~\bibnamefont {Pagonabarraga}}, \bibinfo
  {author} {\bibfnamefont {S.}~\bibnamefont {S{\'a}nchez}}, \ and\ \bibinfo
  {author} {\bibfnamefont {A.}~\bibnamefont {G~Grushin}},\ }\href
  {https://www.nature.com/articles/s41467-021-24948-2} {\bibfield  {journal}
  {\bibinfo  {journal} {Nature Communications}\ }\textbf {\bibinfo {volume}
  {12}},\ \bibinfo {pages} {1} (\bibinfo {year} {2021})}\BibitemShut {NoStop}%
\bibitem [{\citenamefont {Zhang}\ \emph {et~al.}(2021)\citenamefont {Zhang},
  \citenamefont {Tian}, \citenamefont {Jiang}, \citenamefont {Lu},\ and\
  \citenamefont {Chen}}]{zhang2021observation}%
  \BibitemOpen
  \bibfield  {author} {\bibinfo {author} {\bibfnamefont {X.}~\bibnamefont
  {Zhang}}, \bibinfo {author} {\bibfnamefont {Y.}~\bibnamefont {Tian}},
  \bibinfo {author} {\bibfnamefont {J.-H.}\ \bibnamefont {Jiang}}, \bibinfo
  {author} {\bibfnamefont {M.-H.}\ \bibnamefont {Lu}}, \ and\ \bibinfo {author}
  {\bibfnamefont {Y.-F.}\ \bibnamefont {Chen}},\ }\href
  {https://www.nature.com/articles/s41467-021-25716-y} {\bibfield  {journal}
  {\bibinfo  {journal} {Nature communications}\ }\textbf {\bibinfo {volume}
  {12}},\ \bibinfo {pages} {1} (\bibinfo {year} {2021})}\BibitemShut {NoStop}%
\bibitem [{\citenamefont {Liang}\ \emph {et~al.}(2022)\citenamefont {Liang},
  \citenamefont {Xie}, \citenamefont {Dong}, \citenamefont {Li}, \citenamefont
  {Li}, \citenamefont {Gadway}, \citenamefont {Yi},\ and\ \citenamefont
  {Yan}}]{liang2022observation}%
  \BibitemOpen
  \bibfield  {author} {\bibinfo {author} {\bibfnamefont {Q.}~\bibnamefont
  {Liang}}, \bibinfo {author} {\bibfnamefont {D.}~\bibnamefont {Xie}}, \bibinfo
  {author} {\bibfnamefont {Z.}~\bibnamefont {Dong}}, \bibinfo {author}
  {\bibfnamefont {H.}~\bibnamefont {Li}}, \bibinfo {author} {\bibfnamefont
  {H.}~\bibnamefont {Li}}, \bibinfo {author} {\bibfnamefont {B.}~\bibnamefont
  {Gadway}}, \bibinfo {author} {\bibfnamefont {W.}~\bibnamefont {Yi}}, \ and\
  \bibinfo {author} {\bibfnamefont {B.}~\bibnamefont {Yan}},\ }\href
  {https://arxiv.org/abs/2201.09478} {\bibfield  {journal} {\bibinfo  {journal}
  {arXiv preprint arXiv:2201.09478}\ } (\bibinfo {year} {2022})}\BibitemShut
  {NoStop}%
\bibitem [{\citenamefont {Longhi}(2019{\natexlab{a}})}]{longhi2019non}%
  \BibitemOpen
  \bibfield  {author} {\bibinfo {author} {\bibfnamefont {S.}~\bibnamefont
  {Longhi}},\ }\href
  {https://opg.optica.org/ol/abstract.cfm?uri=ol-44-23-5804s} {\bibfield
  {journal} {\bibinfo  {journal} {Optics letters}\ }\textbf {\bibinfo {volume}
  {44}},\ \bibinfo {pages} {5804} (\bibinfo {year}
  {2019}{\natexlab{a}})}\BibitemShut {NoStop}%
\bibitem [{\citenamefont {Longhi}(2019{\natexlab{b}})}]{Longhi19prr}%
  \BibitemOpen
  \bibfield  {author} {\bibinfo {author} {\bibfnamefont {S.}~\bibnamefont
  {Longhi}},\ }\href {\doibase 10.1103/PhysRevResearch.1.023013} {\bibfield
  {journal} {\bibinfo  {journal} {Phys. Rev. Research}\ }\textbf {\bibinfo
  {volume} {1}},\ \bibinfo {pages} {023013} (\bibinfo {year}
  {2019}{\natexlab{b}})}\BibitemShut {NoStop}%
\bibitem [{\citenamefont {Xiao}\ \emph {et~al.}(2021)\citenamefont {Xiao},
  \citenamefont {Deng}, \citenamefont {Wang}, \citenamefont {Wang},
  \citenamefont {Yi},\ and\ \citenamefont {Xue}}]{xiao21prl}%
  \BibitemOpen
  \bibfield  {author} {\bibinfo {author} {\bibfnamefont {L.}~\bibnamefont
  {Xiao}}, \bibinfo {author} {\bibfnamefont {T.}~\bibnamefont {Deng}}, \bibinfo
  {author} {\bibfnamefont {K.}~\bibnamefont {Wang}}, \bibinfo {author}
  {\bibfnamefont {Z.}~\bibnamefont {Wang}}, \bibinfo {author} {\bibfnamefont
  {W.}~\bibnamefont {Yi}}, \ and\ \bibinfo {author} {\bibfnamefont
  {P.}~\bibnamefont {Xue}},\ }\href {\doibase 10.1103/PhysRevLett.126.230402}
  {\bibfield  {journal} {\bibinfo  {journal} {Phys. Rev. Lett.}\ }\textbf
  {\bibinfo {volume} {126}},\ \bibinfo {pages} {230402} (\bibinfo {year}
  {2021})}\BibitemShut {NoStop}%
\bibitem [{\citenamefont {Song}\ \emph {et~al.}(2021)\citenamefont {Song},
  \citenamefont {Wang},\ and\ \citenamefont {Wang}}]{song2021non}%
  \BibitemOpen
  \bibfield  {author} {\bibinfo {author} {\bibfnamefont {F.}~\bibnamefont
  {Song}}, \bibinfo {author} {\bibfnamefont {H.-Y.}\ \bibnamefont {Wang}}, \
  and\ \bibinfo {author} {\bibfnamefont {Z.}~\bibnamefont {Wang}},\ }\href
  {https://arxiv.org/abs/2102.02230} {\bibfield  {journal} {\bibinfo  {journal}
  {arXiv preprint arXiv:2102.02230}\ } (\bibinfo {year} {2021})}\BibitemShut
  {NoStop}%
\bibitem [{\citenamefont {Yi}\ and\ \citenamefont {Yang}(2020)}]{Yi20prl}%
  \BibitemOpen
  \bibfield  {author} {\bibinfo {author} {\bibfnamefont {Y.}~\bibnamefont
  {Yi}}\ and\ \bibinfo {author} {\bibfnamefont {Z.}~\bibnamefont {Yang}},\
  }\href {\doibase 10.1103/PhysRevLett.125.186802} {\bibfield  {journal}
  {\bibinfo  {journal} {Phys. Rev. Lett.}\ }\textbf {\bibinfo {volume} {125}},\
  \bibinfo {pages} {186802} (\bibinfo {year} {2020})}\BibitemShut {NoStop}%
\bibitem [{\citenamefont {Xue}\ \emph {et~al.}(2021)\citenamefont {Xue},
  \citenamefont {Li}, \citenamefont {Hu}, \citenamefont {Song},\ and\
  \citenamefont {Wang}}]{Xue21prb}%
  \BibitemOpen
  \bibfield  {author} {\bibinfo {author} {\bibfnamefont {W.-T.}\ \bibnamefont
  {Xue}}, \bibinfo {author} {\bibfnamefont {M.-R.}\ \bibnamefont {Li}},
  \bibinfo {author} {\bibfnamefont {Y.-M.}\ \bibnamefont {Hu}}, \bibinfo
  {author} {\bibfnamefont {F.}~\bibnamefont {Song}}, \ and\ \bibinfo {author}
  {\bibfnamefont {Z.}~\bibnamefont {Wang}},\ }\href {\doibase
  10.1103/PhysRevB.103.L241408} {\bibfield  {journal} {\bibinfo  {journal}
  {Phys. Rev. B}\ }\textbf {\bibinfo {volume} {103}},\ \bibinfo {pages}
  {L241408} (\bibinfo {year} {2021})}\BibitemShut {NoStop}%
\bibitem [{\citenamefont {Hatano}\ and\ \citenamefont
  {Nelson}(1996)}]{HN96prl}%
  \BibitemOpen
  \bibfield  {author} {\bibinfo {author} {\bibfnamefont {N.}~\bibnamefont
  {Hatano}}\ and\ \bibinfo {author} {\bibfnamefont {D.~R.}\ \bibnamefont
  {Nelson}},\ }\href {\doibase 10.1103/PhysRevLett.77.570} {\bibfield
  {journal} {\bibinfo  {journal} {Phys. Rev. Lett.}\ }\textbf {\bibinfo
  {volume} {77}},\ \bibinfo {pages} {570} (\bibinfo {year} {1996})}\BibitemShut
  {NoStop}%
\bibitem [{\citenamefont {Jiang}\ \emph {et~al.}(2019)\citenamefont {Jiang},
  \citenamefont {Lang}, \citenamefont {Yang}, \citenamefont {Zhu},\ and\
  \citenamefont {Chen}}]{Chenshu19prb}%
  \BibitemOpen
  \bibfield  {author} {\bibinfo {author} {\bibfnamefont {H.}~\bibnamefont
  {Jiang}}, \bibinfo {author} {\bibfnamefont {L.-J.}\ \bibnamefont {Lang}},
  \bibinfo {author} {\bibfnamefont {C.}~\bibnamefont {Yang}}, \bibinfo {author}
  {\bibfnamefont {S.-L.}\ \bibnamefont {Zhu}}, \ and\ \bibinfo {author}
  {\bibfnamefont {S.}~\bibnamefont {Chen}},\ }\href {\doibase
  10.1103/PhysRevB.100.054301} {\bibfield  {journal} {\bibinfo  {journal}
  {Phys. Rev. B}\ }\textbf {\bibinfo {volume} {100}},\ \bibinfo {pages}
  {054301} (\bibinfo {year} {2019})}\BibitemShut {NoStop}%
\bibitem [{\citenamefont {Liu}\ \emph {et~al.}(2021)\citenamefont {Liu},
  \citenamefont {Wang}, \citenamefont {Zheng},\ and\ \citenamefont
  {Chen}}]{Chenshu21prb}%
  \BibitemOpen
  \bibfield  {author} {\bibinfo {author} {\bibfnamefont {Y.}~\bibnamefont
  {Liu}}, \bibinfo {author} {\bibfnamefont {Y.}~\bibnamefont {Wang}}, \bibinfo
  {author} {\bibfnamefont {Z.}~\bibnamefont {Zheng}}, \ and\ \bibinfo {author}
  {\bibfnamefont {S.}~\bibnamefont {Chen}},\ }\href {\doibase
  10.1103/PhysRevB.103.134208} {\bibfield  {journal} {\bibinfo  {journal}
  {Phys. Rev. B}\ }\textbf {\bibinfo {volume} {103}},\ \bibinfo {pages}
  {134208} (\bibinfo {year} {2021})}\BibitemShut {NoStop}%
\bibitem [{\citenamefont {Abo-Shaeer}\ \emph {et~al.}(2001)\citenamefont
  {Abo-Shaeer}, \citenamefont {Raman}, \citenamefont {Vogels},\ and\
  \citenamefont {Ketterle}}]{abo2001observation}%
  \BibitemOpen
  \bibfield  {author} {\bibinfo {author} {\bibfnamefont {J.~R.}\ \bibnamefont
  {Abo-Shaeer}}, \bibinfo {author} {\bibfnamefont {C.}~\bibnamefont {Raman}},
  \bibinfo {author} {\bibfnamefont {J.~M.}\ \bibnamefont {Vogels}}, \ and\
  \bibinfo {author} {\bibfnamefont {W.}~\bibnamefont {Ketterle}},\ }\href
  {https://www.science.org/doi/full/10.1126/science.1060182?casa_token=T7ziefO0eJEAAAAA%3AHs7PC5rP-mCNRihrPkA-K--hQbCzn3tu5MRpm8b1QHIboYeWrHmKV-KFNdQvNK2s8xpdSxsCgbwd8Q}
  {\bibfield  {journal} {\bibinfo  {journal} {Science}\ }\textbf {\bibinfo
  {volume} {292}},\ \bibinfo {pages} {476} (\bibinfo {year}
  {2001})}\BibitemShut {NoStop}%
\bibitem [{\citenamefont {Lin}\ \emph {et~al.}(2009)\citenamefont {Lin},
  \citenamefont {Compton}, \citenamefont {Jim{\'e}nez-Garc{\'\i}a},
  \citenamefont {Porto},\ and\ \citenamefont {Spielman}}]{lin2009synthetic}%
  \BibitemOpen
  \bibfield  {author} {\bibinfo {author} {\bibfnamefont {Y.-J.}\ \bibnamefont
  {Lin}}, \bibinfo {author} {\bibfnamefont {R.~L.}\ \bibnamefont {Compton}},
  \bibinfo {author} {\bibfnamefont {K.}~\bibnamefont
  {Jim{\'e}nez-Garc{\'\i}a}}, \bibinfo {author} {\bibfnamefont {J.~V.}\
  \bibnamefont {Porto}}, \ and\ \bibinfo {author} {\bibfnamefont {I.~B.}\
  \bibnamefont {Spielman}},\ }\href
  {https://www.nature.com/articles/nature08609} {\bibfield  {journal} {\bibinfo
   {journal} {Nature}\ }\textbf {\bibinfo {volume} {462}},\ \bibinfo {pages}
  {628} (\bibinfo {year} {2009})}\BibitemShut {NoStop}%
\bibitem [{\citenamefont {Zhang}\ and\ \citenamefont
  {Franz}(2020)}]{Zhang20prl}%
  \BibitemOpen
  \bibfield  {author} {\bibinfo {author} {\bibfnamefont {X.-X.}\ \bibnamefont
  {Zhang}}\ and\ \bibinfo {author} {\bibfnamefont {M.}~\bibnamefont {Franz}},\
  }\href {\doibase 10.1103/PhysRevLett.124.046401} {\bibfield  {journal}
  {\bibinfo  {journal} {Phys. Rev. Lett.}\ }\textbf {\bibinfo {volume} {124}},\
  \bibinfo {pages} {046401} (\bibinfo {year} {2020})}\BibitemShut {NoStop}%
\bibitem [{\citenamefont {Lin}\ \emph {et~al.}(2021)\citenamefont {Lin},
  \citenamefont {Ding}, \citenamefont {Ke},\ and\ \citenamefont
  {Li}}]{lin2021steeringOL}%
  \BibitemOpen
  \bibfield  {author} {\bibinfo {author} {\bibfnamefont {Z.}~\bibnamefont
  {Lin}}, \bibinfo {author} {\bibfnamefont {L.}~\bibnamefont {Ding}}, \bibinfo
  {author} {\bibfnamefont {S.}~\bibnamefont {Ke}}, \ and\ \bibinfo {author}
  {\bibfnamefont {X.}~\bibnamefont {Li}},\ }\href
  {https://opg.optica.org/ol/abstract.cfm?uri=ol-46-15-3512} {\bibfield
  {journal} {\bibinfo  {journal} {Optics Letters}\ }\textbf {\bibinfo {volume}
  {46}},\ \bibinfo {pages} {3512} (\bibinfo {year} {2021})}\BibitemShut
  {NoStop}%
\bibitem [{\citenamefont {Harper}(1955)}]{harper1955single}%
  \BibitemOpen
  \bibfield  {author} {\bibinfo {author} {\bibfnamefont {P.~G.}\ \bibnamefont
  {Harper}},\ }\href
  {https://iopscience.iop.org/article/10.1088/0370-1298/68/10/304} {\bibfield
  {journal} {\bibinfo  {journal} {Proceedings of the Physical Society. Section
  A}\ }\textbf {\bibinfo {volume} {68}},\ \bibinfo {pages} {874} (\bibinfo
  {year} {1955})}\BibitemShut {NoStop}%
\bibitem [{\citenamefont {Hofstadter}(1976)}]{Hofstadter1976prb}%
  \BibitemOpen
  \bibfield  {author} {\bibinfo {author} {\bibfnamefont {D.~R.}\ \bibnamefont
  {Hofstadter}},\ }\href {\doibase 10.1103/PhysRevB.14.2239} {\bibfield
  {journal} {\bibinfo  {journal} {Phys. Rev. B}\ }\textbf {\bibinfo {volume}
  {14}},\ \bibinfo {pages} {2239} (\bibinfo {year} {1976})}\BibitemShut
  {NoStop}%
\bibitem [{sm()}]{sm}%
  \BibitemOpen
  \href@noop {} {}\bibinfo {note} {See Supplemental Material for the
  suppression of the NHSE effect, scaling of the low-energy Landau levels,
  derivation of the semiclassical orbits and real Landau levels determined by
  the quantization conditions, magnetic spectra under the $x$-OBC, the proof of
  the exponential phase boundary, the size effect of the $\mathcal{MT}$ phase
  transition, $\mathcal{MT}$-breaking of the Hamiltonian by non-Hermitian
  complex hopping and physical results for the nonreciprocal honeycomb lattice,
  which includes Ref.~\cite{Guo21prl}.}\BibitemShut {Stop}%
\bibitem [{\citenamefont {Lu}\ \emph {et~al.}(2021)\citenamefont {Lu},
  \citenamefont {Zhang},\ and\ \citenamefont {Franz}}]{lu2021magnetic}%
  \BibitemOpen
  \bibfield  {author} {\bibinfo {author} {\bibfnamefont {M.}~\bibnamefont
  {Lu}}, \bibinfo {author} {\bibfnamefont {X.-X.}\ \bibnamefont {Zhang}}, \
  and\ \bibinfo {author} {\bibfnamefont {M.}~\bibnamefont {Franz}},\ }\href
  {\doibase 10.1103/PhysRevLett.127.256402} {\bibfield  {journal} {\bibinfo
  {journal} {Phys. Rev. Lett.}\ }\textbf {\bibinfo {volume} {127}},\ \bibinfo
  {pages} {256402} (\bibinfo {year} {2021})}\BibitemShut {NoStop}%
\bibitem [{\citenamefont {Onsager}(1952)}]{onsager1952interpretation}%
  \BibitemOpen
  \bibfield  {author} {\bibinfo {author} {\bibfnamefont {L.}~\bibnamefont
  {Onsager}},\ }\href@noop {} {\bibfield  {journal} {\bibinfo  {journal} {The
  London, Edinburgh, and Dublin Philosophical Magazine and Journal of Science}\
  }\textbf {\bibinfo {volume} {43}},\ \bibinfo {pages} {1006} (\bibinfo {year}
  {1952})}\BibitemShut {NoStop}%
\bibitem [{\citenamefont {Lifshitz}\ and\ \citenamefont
  {Kosevich}(1954)}]{lifshitz1954}%
  \BibitemOpen
  \bibfield  {author} {\bibinfo {author} {\bibfnamefont {I.}~\bibnamefont
  {Lifshitz}}\ and\ \bibinfo {author} {\bibfnamefont {A.}~\bibnamefont
  {Kosevich}},\ }\href@noop {} {\bibfield  {journal} {\bibinfo  {journal}
  {Dokl. Akad. Nauk SSSR}\ }\textbf {\bibinfo {volume} {96}},\ \bibinfo {pages}
  {963} (\bibinfo {year} {1954})}\BibitemShut {NoStop}%
\bibitem [{\citenamefont {Wigner}(1993)}]{wigner1993normal}%
  \BibitemOpen
  \bibfield  {author} {\bibinfo {author} {\bibfnamefont {E.~P.}\ \bibnamefont
  {Wigner}},\ }in\ \href
  {https://link.springer.com/chapter/10.1007/978-3-662-02781-3_38} {\emph
  {\bibinfo {booktitle} {The Collected Works of Eugene Paul Wigner}}}\
  (\bibinfo  {publisher} {Springer},\ \bibinfo {year} {1993})\ pp.\ \bibinfo
  {pages} {551--555}\BibitemShut {NoStop}%
\bibitem [{\citenamefont {Shapere}\ and\ \citenamefont
  {Wilczek}(1989)}]{shapere1989geometric}%
  \BibitemOpen
  \bibfield  {author} {\bibinfo {author} {\bibfnamefont {A.}~\bibnamefont
  {Shapere}}\ and\ \bibinfo {author} {\bibfnamefont {F.}~\bibnamefont
  {Wilczek}},\ }\href@noop {} {\emph {\bibinfo {title} {Geometric phases in
  physics}}},\ Vol.~\bibinfo {volume} {5}\ (\bibinfo  {publisher} {World
  scientific},\ \bibinfo {year} {1989})\BibitemShut {NoStop}%
\bibitem [{\citenamefont {Guo}\ \emph {et~al.}(2021)\citenamefont {Guo},
  \citenamefont {Liu}, \citenamefont {Zhao}, \citenamefont {Liu},\ and\
  \citenamefont {Chen}}]{Guo21prl}%
  \BibitemOpen
  \bibfield  {author} {\bibinfo {author} {\bibfnamefont {C.-X.}\ \bibnamefont
  {Guo}}, \bibinfo {author} {\bibfnamefont {C.-H.}\ \bibnamefont {Liu}},
  \bibinfo {author} {\bibfnamefont {X.-M.}\ \bibnamefont {Zhao}}, \bibinfo
  {author} {\bibfnamefont {Y.}~\bibnamefont {Liu}}, \ and\ \bibinfo {author}
  {\bibfnamefont {S.}~\bibnamefont {Chen}},\ }\href {\doibase
  10.1103/PhysRevLett.127.116801} {\bibfield  {journal} {\bibinfo  {journal}
  {Phys. Rev. Lett.}\ }\textbf {\bibinfo {volume} {127}},\ \bibinfo {pages}
  {116801} (\bibinfo {year} {2021})}\BibitemShut {NoStop}%
\end{thebibliography}
\end{document}